\documentclass[nofootinbib, reprint, amsmath,amssymb, aps,]{revtex4-1}
\usepackage{graphicx}% Include figure files
\usepackage{dcolumn}% Align table columns on decimal point
\usepackage{bm}% bold math
\usepackage[colorlinks]{hyperref}

\usepackage{graphicx,times}
\usepackage{subfigure}
\usepackage{amsmath}
\usepackage{mathrsfs}
\usepackage{cases}
\usepackage{bm}
\usepackage{longtable}
\usepackage{hyperref}
\usepackage{epstopdf}
\usepackage{amsmath,bm}
\usepackage{amssymb}
\usepackage{natbib}
\usepackage{morefloats}
\usepackage{multirow}
\usepackage{array}
\usepackage{verbatim}
\usepackage[stable]{footmisc}
\usepackage{stackrel}
\usepackage{graphicx}
\usepackage{dcolumn}
\usepackage{bm}

\newcommand{\br}{{\bf r}}

\newcommand{\bx}{{\bf x}}

\begin{document}

%\preprint{APS/123-QED}

%\title{Magnetic Topology and Field-Fluid Slippage}
\title{Electromagnetic Helicity in Classical Physics}

\author{Amir Jafari}
 \email{elenceq@jhu.edu}
% \altaffiliation {Department of Physics \& Astronomy, Johns Hopkins University, Baltimore, MD, USA}%Lines break automatically or can be forced with \\

% \altaffiliation {Department of Physics \& Astronomy, Johns Hopkins University, Baltimore, MD, USA}
\affiliation{%
Johns Hopkins University, Baltimore, USA\\
% This line break forced with \textbackslash\textbackslash
}%

\iffalse
\collaboration{MUSO Collaboration}%\noaffiliation

\author{Charlie Author}
 \homepage{http://www.Second.institution.edu/~Charlie.Author}
\affiliation{
 Second institution and/or address\\
 This line break forced% with \\
}%
\affiliation{
 Third institution, the second for Charlie Author
}%
\author{Delta Author}
\affiliation{%
 Authors' institution and/or address\\
 This line break forced with \textbackslash\textbackslash
}%

\collaboration{CLEO Collaboration}%\noaffiliation
\fi
%\date{\today}% It is always \today, today,
             %  but any date may be explicitly specified

\begin{abstract}
This pedagogical note revisits the concept of electromagnetic helicity in classical systems. In particular, magnetic helicity and its role in mean field dynamo theories is briefly discussed highlighting the major mathematical inconsistency in most of these theories---violation of magnetic helicity conservation. A short review of kinematic dynamo theory and its classic theorems is also presented in the Appendix.
\end{abstract}

\pacs{Valid PACS appear here}% PACS, the Physics and Astronomy
                             % Classification Scheme.
%\keywords{Suggested keywords}%Use showkeys class option if keyword
                              %display desired
\maketitle

%\tableofcontents

\section{\label{sec:level1}Introduction}

In classical physics, helicity is usually defined as the inner product of two vector fields integrated over a volume. An example is the cross helicity; scalar product of magnetic and velocity fields integrated over a volume in an electrically conducting fluid. If one of the vector fields is the curl of the other, then helicity measures the structural complexity---twistedness and knottedness---of the curl field. For instance, magnetic helicity $\int_V {\bf A.B} d^3x$ measures the twistedness and knottedness of the magnetic field $\bf B=\nabla\times A$. Magnetic helicity is conserved in ideal magnetohydrodynamics (MHD) similar, in some ways, to kinetic helicity which is conserved in inviscid fluids. Even in magnetically diffusive media, magnetic helicity is still better conserved than the energy. For example, unlike magnetic energy, the helicity is often assumed to be well conserved in the process of magnetic reconnection--eruptive fluid motions caused by an evolving magnetic field (for a review of reconnection see e.g., \cite{JafariandVishniac2018}; \cite{Jafarietal2018} and references therein. A more recent picture which considers magnetic topology and stochasticity can be found in \cite{JV2019}). This gives magnetic helicity enormous importance in a variety of problems. Dynamo theory is an example, perhaps the most important one, in which magnetic helicity has been realized to play a very crucial role.

The integrand in the definition of magnetic helicity, called the helicity density, obeys a continuity-like equation. In fact, as we will see in detail, the magnetic helicity four-vector

$${J}_M^\mu=({\bf A.B}, \bf{B}\phi+\bf{E\times A})$$

has a vanishing divergence in ideal MHD, i.e., $\partial_\mu{J}_M^\mu=0$, similar to current four-vector ${ j}^\mu=(\rho, {\bf j})$ in electromagnetism; $\partial_\mu{j}^\mu=0$. Here $\phi$, $\bf E$, $\rho$ and $\bf j$ denote, respectively, the electric scalar potential, electric field, charge density and electric current. One might speculate about a deeper underlying physics behind the helicity equation similar to the gauge independence in electromagnetism which is closely related to the charge conservation. Indeed, this situation resembles many other in, for instance, quantum field theories and particle physics. This provides enough motivation to look at magnetic helicity from a field-theory point of view. This realization would become more vivid when one tries to formulate MHD using a Lagrangian formalism. Obviously, the helicity equation is covariant under Lorentz transformations and its form is independent of gauge. However, the helicity four-vector is gauge dependent and are obviously so both its timelike component, helicity density, and spacelike component, helicity flux. The gauge dependence of helicity density has motivated several researchers to re-define it in slightly different ways to circumvent the gauge issue. Nevertheless, one might look at helicity as a concept similar to potential. No matter what potential is chosen for the ground level of a system, the potential difference between two points retains its physical meaning. Once we fix a gauge, we can talk about the evolution of magnetic helicity and its conservation without any concern about helicity's gauge dependence. The important point is that, with any gauge chosen, helicity four-vector can describe gauge-independent physical phenomena. For instance, its spacelike component, helicity flux, is closely related to the $\alpha$-effect in dynamo theories. Detailed discussions of magnetic helicity and its applications in different contexts can be found e.g., in \cite{Berger1984}; \cite{Moffattetal1992}; \cite{Berger1999}; \cite{Brandenburg2009}; \cite{Blackman2015}. 			
	
In Section II, we formulate magnetic helicity mathematically emphasizing its relativistic four-vector formalism. Section III is a brief review of mean field dynamo theories highlighting the role magnetic helicity plays in them. We will set the speed of light to unity, $c=1$, and also adopt the metric signature $(-+++)$ for the Minkowski space-time.

\section{Concept of Helicity}

A pack of spaghetti sitting on the shelf of a supermarket consists of almost parallel straight lines with no twisted or entangled form. In a hot dish on the dinner table, however, it is in fact swirls of entangled and twisted threads of spaghetti which resemble curves in the 3-dimensional Euclidean space. Mathematically, we can define a "linking number" between one pair of curves in three dimensions as a measure of their swirl and twistedness around each other. We may look for available vectors to make a scalar as the linking number associated with such pair of curves. Consider point A on one curve, $\vec{x}(\sigma)$, and point B on the other one, $\vec{y}(\tau)$, with $\sigma$ and $\tau$ being arbitrary curve parameterizations. The tangent vectors at these points plus the distance vector connecting the two, $\vec{r}=\vec{x}-\vec{y}$, are three appropriate vectors. We can obtain a scalar function by making an inner product between one of these vectors and the cross product of the other two. Dividing by the cube of the distance between the two points, $r^3$, we get a dimensionless scalar function. Integrating this scalar over the curves defines the Gauss linking number. For two parametric curves, $\vec{x}(\sigma)$ and $\vec{y}(\tau)$, we write
\begin{equation}\label{1}
\oint_\sigma \oint_\tau \Big({d\vec{x}\over d\sigma}\times {d\vec{y}\over d\tau}\Big).{\vec{x}-\vec{y}\over |\vec{x}-\vec{y}| ^3 } d\sigma d\tau.
\end{equation}
Next, we replace the curves with magnetic flux tubes and sum the linking numbers between all pairs of such flux tubes. Before doing this, and in order to get an insight into magnetic flux tubes, consider the time evolution of the magnetic flux, $\Phi$, of a closed curve $C$ bounded by the surface $S$ in a non-diffusive medium with velocity field $\mathbf{v}$;

\begin{equation}\notag
{\partial\over\partial t}\int_S \mathbf{B.}d\mathbf{S}=\int_S{\partial\mathbf{B}\over\partial t} .d\mathbf{S}-\oint_C\mathbf{v\times B.}d\mathbf{l},
\end{equation}
which upon using the induction equation, $\partial_t\mathbf{B}=\nabla\times(\mathbf{v}\times\mathbf{B})$, leads to the Alfv\'en flux freezing theorem: the flux through a surface surrounded by a closed curve co-moving with the fluid is constant. Sweeping the boundary curve $C$ along the local magnetic field defines a magnetic flux tube.  In the limit when the diamater of the tube tends to zero \cite{Biskamp1997} we will have a magnetic field line which can also be defined by the integral curves of $\mathbf{B}\times d\mathbf{x}=0$. Back to eq.(\ref{1}), we note that the magnetic field of any flux tube points in the direction of the local tangent vector and so we can define magnetic helicity of the two flux tubes $\mathbf{B}(\vec{x})$ and $\mathbf{B}(\vec{y})$ as
\begin{equation}\label{helicity0}
{\cal{J}}_M={-1\over 4\pi} \int_V \int_V d^3x d^3y \Big( \mathbf{B}(\vec{x})\times \mathbf{B}(\vec{y})\Big) .{\vec{r}\over r^3 }.
\end{equation}
One can obtain an even more compact expression by re-arranging the triple vector product and using the Coulomb gauge vector potential
\begin{equation}
\mathbf{A}(\vec{x})={1\over 4\pi} \int_V   \mathbf{B}(\vec{y}) \times{\vec{r}\over r^3 }d^3 y.
\end{equation}
The result is
\begin{equation}\label{helicity10}
{\cal{J}}_M= \int_V  \mathbf{A.}\mathbf{B} d^3x  .
\end{equation}
Since $\nabla\times\mathbf{A}$ measures the rotation of the vector field $\mathbf{A}$, so its inner product with $\mathbf{A}$ is an indicator of how much this vector field rotates around itself, hence the name helicity. The integrand in eq.(\ref{helicity10}) is called the magnetic helicity density or even sometimes simply magnetic helicity $J^0_M=\mathbf{A.}\mathbf{B} $. In a similar manner, one may define electric helicity $J^0_E=\mathbf{E.C}$ where the electric potential $\mathbf{C}$ is defined through $\nabla\times\mathbf{C}=\mathbf{E}$ when $\nabla.\mathbf{j}=0$. These are, as we will see later on, the time components of two corresponding four-vectors; the magnetic helicity four-vector ${J}^\mu_M=(J^0_M,\mathbf{J}_M)$ and the electric helicity four-vector ${J}^\mu_E=(J^0_E,\mathbf{J}_E)$. In terms of the electromagnetic field tensor  $F_{\mu\nu}=\partial_\mu A_\nu-\partial_\nu A_\mu$ and its dual $G^{\mu\nu}$, we have
\begin{equation}
J_M^\mu=-A_\nu G^{\mu\nu}, 
\end{equation}
and
\begin{equation}
J_E^\mu=C_\nu F^{\mu\nu}.
\end{equation}
The electromagnetic helicity is the sum of these two quantities; 
\begin{equation}\label{EM-helicity}
J^\mu_{EM}=C_\nu F^{\mu\nu}-A_\nu G^{\mu\nu}.
\end{equation}

It turns out that the electromagnetic helicity is closely related to the particle helicity in high energy physics (projection of the angular momentum onto the momentum of a particle). In fact, it represents the difference between the numbers of right-handed and left-handed photons \cite{Truebaetal1996}. We should emphasize, however, that the above definition of electric helicity does not hold in MHD since in the latter case, the electric field cannot generally be expressed as a curl of some potential field. The other way to define electric helicity is to write it as $A_\nu F^{\mu\nu}$. We will not pursue this procedure here and instead will focus on the magnetic helicity in MHD. The only exception will be mentioning some analogous expressions for $J_E^\mu=C_\nu F^{\mu\nu}$ to show the mathematical similarities with electromagnetism.

\begin{figure}
\includegraphics[scale=.6]{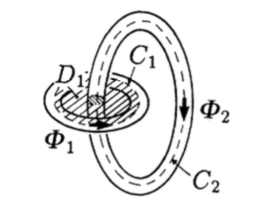}
\centering
\caption {\footnotesize {Should be re-drawn: An illustration of two linked but not knotted flux tubes oriented along two curves $C_1$ and $C_2$ carrying magnetic fluxes $\Phi_1$ and $\Phi_2$. The corresponding helicity is $2\Phi_1\Phi_2$. Illustration from \cite{Moffattetal1992}.}}
\label{Moffatt-Ricca1992}
\end{figure}

\subsection{Evolution Equation}

\begin{figure}
\includegraphics[scale=.4]{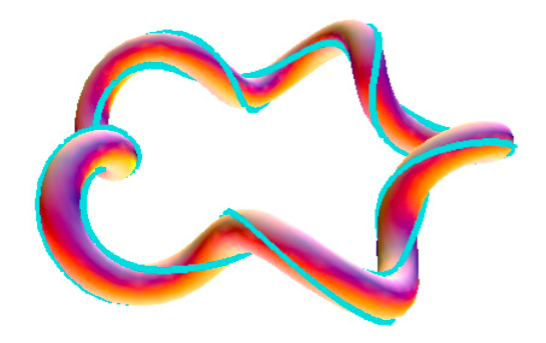}
\centering
\caption {\footnotesize {A magnetic flux tube with twist (around the magnetic axis) and writhe (winding of the tube's axis itself). The net helicity of an individual flux tube is the sum of its writhe and twist. Illustration from \cite{Berger1999}.}}
\label{Berger1999}
\end{figure}

Consider two linked magnetic flux tubes as shown in Fig.(\ref{Moffatt-Ricca1992}). The total magnetic helicity is ${\cal{J}}_M=\int_{V_1} \mathbf{A.B}+\int_{V_2} \mathbf{A.B}$. Rearranging the integrals and using $d^3x\rightarrow d\mathbf{S.}d\mathbf{x}$, each of these integrals can be written as
\begin{equation}\notag
\oint_{S_1} \oint_{C_1} \mathbf{A.B} (d\mathbf{S.}d\mathbf{x})=\oint_{S_1} \oint_{C_1} \mathbf{B.} d\mathbf{S}\mathbf{A.}d\mathbf{x}=\Phi_1\Phi_2,
\end{equation}
where $\Phi_1$ and $\Phi_2$ are magnetic fluxes in the tubes $C_1$ and $C_2$. The total helicity is therefore ${\cal{J}}_M=2\Phi_1\Phi_2$. This simple consideration can be immediately extended to $N$ flux tubes (in an arbitrary configuration) each carrying magnetic flux $\Phi_i$;
\begin{equation}
{\cal{J}}_M=\sum_{i=1}^{N} \sum_{j=1}^{N}L_{ij}\Phi_i\Phi_j,
\end{equation}
where $L_{ij}$ are constant coefficients. In the limit, when $N\rightarrow \infty$ and $\Phi_i\rightarrow 0$, we recover eq.(\ref{helicity0}). Following Berger \cite{Berger1999}, we call each of the $N(N-1)$ terms in the above sum in which $i\neq j$ having the form $2L_{ij}\Phi_i\Phi_j$, the mutual helicity. The terms with $i=j$, $L_{ii}\Phi_i^2$, represent self-helicities. The coefficient $L_{ii}$ in the self-helicity terms corresponds to the internal twists of the flux tubes. The mutual helicity measures linking between the tubes whereas the self-helicity is defined for an individual tube as the sum of its twist and writhe: writhe is the linking and kinking of the magnetic field axis of the tube; see Fig.(\ref{Berger1999}). For a detailed discussion about the relationship between twist, writhe and linkage (see e.g., \cite{Blackman2015}).

Magnetic helicity, as a robustly conserved quantity in ideal MHD, is also the key to understanding the inverse cascade in turbulent MHD systems. Even in systems with zero total helicity, one can still get interesting effects if the system generates a separation of positive and negative helicities.
In addition, helicity four-vector is the only classical example of a Cherns-Simons symmetry. Gauge freedom in this context means that this is actually a family of conservation laws, although not all are physically interesting. As far as the topology of the vector fields is concerned, magnetic helicity is one special case of a more general concept. The helicity corresponding to a divergence-free vector field on a 3-dimensional manifold measures the twists of the associated integral lines of the field \cite{Woltjer1958}. So, for instance, kinetic and magnetic helicities are measures of twists and linkages of, respectively, vorticity and magnetic field lines. To give a general definition, suppose that $\mathbf{F}(x)$, $x\in \cal{M}$, is a real divergenceless vector field  (i.e., a solenoidal field; $\nabla.\mathbf{F}=0$) in the $3$-dimensional parallelizable manifold $\cal{M}$;

$$F: {\cal{M}}\rightarrow R^3.$$

There exists a vector potential field $\mathbf{W}$ in $\cal{M}$ satisfying $\mathbf{F}=\nabla\times \mathbf{W}$. Depending on the cohomology of the manifold $\cal{M}$ the definition of this vector potential may be local or global. The helicity associated with a solenoidal vector field $\mathbf{F}$ is defined as $\int d^3x\mathbf{F.} curl^{-1} \mathbf{F}$. This is the topological linkage between the tubes in which $\mathbf{F}$ is non-zero. In general, calculation of $curl^{-1} \mathbf{F}$ can be problematic because $curl^{-1}\mathbf{F}$ can be replaced with $curl^{-1}\mathbf{F}+\nabla\phi$ without changing the helicity. Nevertheless, this gauge dependence becomes a problem only for vector fields whose inverse curl is not a "measurable" quantity. If the inverse curl of $\mathbf{F}$ carries a measurable physical concept, then there is no difficulty associated with the gauge. For example, the inverse curl of the vorticity $\mathbf{w} $ is the velocity field which is physically measurable. Therefore, for the kinetic helicity, the question of gauge dependence never arises. On the other hand, in the case of magnetic helicity, the inverse curl is the vector potential which is not physically measurable. Also, it is important to note that in the case of periodic boundary conditions, or when $\mathbf{F.} \mathbf{n}=0$ on the boundaries where $ \mathbf{n}$ is normal vector on the boundary, the helicity becomes gauge-invariant. This can easily be seen writing
\begin{eqnarray}\notag
\int_V d^3x \mathbf{F.}\Big( curl^{-1} \mathbf{F}+&&\nabla \phi\Big)=\\\notag
&&\int_V d^3x\mathbf{F.}curl^{-1}\mathbf{F}+\int_V d^3x\mathbf{F.}\nabla\phi.
\end{eqnarray}
The second integral at the RHS of the above equation can be converted to a surface integral which, by virtue of the imposed boundary conditions, vanishes;
\begin{equation}\notag
\int_V d^3x\;\mathbf{F.}{\nabla}\phi= \int_S \phi\mathbf{F.}\mathbf{n} \; dS=0 ,
\end{equation}
where we have used ${\nabla.}\mathbf{F}=0$. In fact, for many other boundary conditions, the magnetic helicity would be gauge dependent.

There is another quantity closely related to the notion of magnetic helicity. In constructing the Lagrangian for the Maxwell's equations, the Lorentz invariant scalar $\mathbf{E.B}$ is usually left out in favor of the other invariant $E^2-B^2=-F^{\mu\nu}F_{\mu\nu}/2$ since the former, unlike the latter, is a pseudo-scalar. However, it turns out that it has a close relationship with the time evolution of magnetic helicity density. To see this, suffice it to express $\mathbf{E.B}$ in terms of potentials using $\mathbf{E}=-\nabla \phi-\partial_t \mathbf{A}$ and $\mathbf{B}=\nabla\times \mathbf{A}$. The result is the helicity equation:
\begin{equation}\label{helicity-1}
-2\mathbf{E.B}={\partial J^0_M\over\partial t}+\nabla.\mathbf{J}_M.
\end{equation}
The divergence term contains the so-called magnetic helicity flux $\mathbf{J}_M=\mathbf{B}\phi+\mathbf{E\times A}$. Using the Ohm's law, we can expand this flux as

\begin{equation}\label{helicity30}
\mathbf{J}_M= \underbrace{ ( \mathbf{A.B}) \mathbf{v} }_\text{Advective Flux}+\underbrace{  \mathbf{B}(\phi-\mathbf{A.v}) }_\text{Dynamical Flux}+\underbrace{ \eta(\mathbf{j} \times \mathbf{A})}_\text{Resistive Flux}.
\end{equation}

Note that with any gauge choice, ${J}^\mu_M=(J^0_M, \mathbf{J}_M)$, or more explicitly $J^\mu_M=(\mathbf{A.B}, \mathbf{B}\phi+\mathbf{E\times A})$, is a four-vector. Therefore, we can write eq.(\ref{helicity-1}) in a covariant form:
\begin{equation}
F^{\mu\nu}G_{\mu\nu}=-2\partial_\mu{J}^\mu_M.
\end{equation}
(A similar expression for the electric helicity is $F^{\mu\nu}G_{\mu\nu}=2\partial_\mu{J}^\mu_E$ when $\nabla.\mathbf{j}=0$). To express this equation differently, one may also define the electric and magnetic four-vectors as
\begin{equation}
{E}^\mu=\gamma(\mathbf{v.}\mathbf{E}, \mathbf{E}+\mathbf{v\times B}),
\end{equation}
and
\begin{equation}
{B}^\mu=\gamma(\mathbf{v.}\mathbf{B}, \mathbf{B}-\mathbf{v\times E}),
\end{equation}
where $\gamma=(1-v^2)^{-1/2}$ is the Lorentz factor and $\mathbf{v}$ is the velocity with four-vector ${U}^\mu=\gamma(1, \mathbf{v})$. These four-vectors also satisfy ${E^\mu U_\mu}={B^\mu U_\mu}=0$. It is easy to show that $\mathbf{E.B}={E^\mu}{B}_\mu$ which is also equal to $-F^{\mu\nu}G_{\mu\nu}/4$. Now, we may express the helicity equation as
\begin{equation}\label{helicity1}
-{1\over 2}F^{\mu\nu}G_{\mu\nu}=-2{E^\mu}{B}_\mu=\partial_\mu {J}^\mu_M.
\end{equation}
(Similarly, for the electric helicity we have $F^{\mu\nu}G_{\mu\nu}/2=2{E^\mu}{B}_\mu=\partial_\mu {J}^\mu_E$ if $\nabla.\mathbf{j}=0$). One may also divide up the magnetic helicity flux $\mathbf{J}_M$ into longitudinal (curl free), $\mathbf{J_l}_M$, and transverse (divergence free), $\mathbf{J_t}_M$, parts. A simple calculation then yields
\begin{equation}
\mathbf{J_l}_M=\nabla \int_V  {\partial J_M^0(\vec{x}')\over\partial t}{d^3x'\over |\vec{x}-\vec{x}'|},
\end{equation}
and
\begin{equation}
\mathbf{J_t}_M=\nabla \times\nabla\times \int_V \mathbf{J}_M(\vec{x}'){d^3x'\over |\vec{x}-\vec{x}'|}.
\end{equation}

 Interestingly, the term $\mathbf{E.B}=E_\mu B^\mu=F_{\mu\nu}G^{\mu\nu}/2$ is a pseudo-scalar Lorentz invariant related to a topological concept which is a gauge dependent physical quantity. In magnetohydrodynamics (MHD), we invoke an additional constraint---the Ohm's law $\mathbf{E}+\mathbf{v\times B}=\eta \mathbf{j}$--- and the current helicity, $\mathbf{j.B}$, becomes part of a source term for the magnetic helicity;
\begin{equation}\label{helicity3}
{\partial J^0_M\over\partial t}+\nabla\mathbf{.J}_M=-2\eta\mathbf{j}\mathbf{.B}.
\end{equation}

For ideal MHD\footnote{If turbulence is present, then the velocity and magnetic fields will not be Lipschitz continuous, consequently their spatial derivatives will be ill-defined in general. A method of renormalization, or coarse-graining, can be applied using the mathematical concept of distributions to obtain well-defined fields; see e.g., \cite{JV2019}.}, we find
\begin{equation}\label{helicity20}
\partial_\mu {J}^\mu_M=0,
\end{equation}
or,
\begin{equation}\notag
{\partial \mathbf{A.B}\over\partial t}+\nabla.\Big((\mathbf{A.B})\mathbf{v}+\mathbf{B}(\phi-\mathbf{A.v})  \Big)=0.
\end{equation}
which represents the conservation of magnetic helicity. (There is an expression similar to (\ref{helicity20}), and another one similar to (\ref{helicity3}) with a flipped sign for current helicity, for the electric helicity provided that $\nabla.\mathbf{j}=0$.) If we take the average of eq.(\ref{helicity3}), by integrating it over a periodic volume or inside a magnetic surface, the divergence term would not contribute. Note that, for periodic boundary conditions, ${\cal{J}}_M$ is invariant meaning that magnetic helicity is physically meaningful \cite{Moffatt1969}. In this case we have
\begin{equation}\label{evolution of magnetic helicity}
{\partial{\cal{J}}_M\over \partial t}={\partial\over \partial t} \int_V \mathbf{A.}\mathbf{B} d^3x= -2 \eta \int_V \mathbf{j.B}d^3x.
\end{equation}
Using the Schwarz's inequality, we find
\begin{equation}
\Big|{\partial {\cal{J}}_M\over \partial t}\Big|\leq 2\sqrt{\eta \Big|{\partial u_B^2\over\partial t} \Big| },
\end{equation}
where $u_B=\int (B^2/2) d^3x$ is the magnetic energy with the dissipation rate
\begin{equation}
\Big| {\partial u_B\over \partial t}  \Big| =\int_V \eta j^2 d^3x.
\end{equation}
Now, we can define an Ohmic dissipation rate as $\tau_d^{-1}=\eta/L^2$ where $L=|{\cal{J}}_M|/u_B$ \cite{Berger1999}. The rate at which the magnetic helicity changes is proportional to a term that vanishes in the limit of large magnetic Reynolds numbers. In other words, magnetic helicity is better conserved than energy and when the resistivity approaches zero, magnetic helicity becomes a perfectly conserved quantity. This fact indeed is in contrast with the case of kinetic helicity, which is also conserved in completely inviscid fluids, whose dissipation at small scales does not vanish as we take the limit when viscosity goes to zero. The latter means that the kinetic helicity is not even approximately conserved in real turbulent flows \cite{VishniacandCho2001}. One can define the spectra of magnetic energy, $E(k)$, and magnetic helicity, $J_M(k)$, using Fourier transforms;
\begin{equation}\label{Fourier for magnetic energy-1}
\int_0^{\infty} u_B(k)dk=\langle {B^2\over 2} \rangle,
\end{equation}
and
\begin{equation}\label{Fourier for magnetic energy-2}
\int_0^\infty J_M(k)dk=\langle \mathbf{A.B}\rangle,
\end{equation}
where $0<k<\infty$ is the wavenumber, and the brackets denote volume averages over a periodic domain. These definitions, using the Schwartz inequality, yield
\begin{equation}\label{Fourier for magnetic energy-2}
{k\over 2}|J_M(k)| \leq u_B(k).
\end{equation}
This is called the realizability condition. The equality holds for the fully helical magnetic fields with positive helicity. In this particular case, the energy and magnetic helicity cannot cascade directly. Therefore, the interaction of modes with wavenumbers $p$ and $q$ can only produce fields whose wavevector $\mathbf{k} $ satisfies $|\mathbf{k}|\leq max \left( |\mathbf{p}|+|\mathbf{q}|\right)$. As a consequence,the magnetic helicity and magnetic energy are transformed to progressively larger length scales \cite{Brandenburg2009}. 

\subsection{Taylor Relaxation and Woltjer Theorem}

At this point, one may ask that with the constraint of helicity conservation, how a magnetized plasma inside a magnetic surface evolves to its minimum energy state. Mathematically, we want to minimize the functional $\int_V u_B$ subject to the condition that magnetic helicity ${\cal{J}}_M$ remains constant. So, introducing a Lagrange multiplier $\lambda/2$, we vary the functional $\int_V u_B-\lambda {\cal{J}}_M/2$. The associated Lagrangian is
\begin{equation}
{\cal{L}}={1\over 2}\Big( \epsilon_{ijk}\partial_iA_j\Big)^2-{\lambda\over 2}\epsilon_{ijk}A_k\partial_iA_j.
\end{equation}
Consequently, the Euler-Lagrange equations
\begin{equation}\notag
{\partial {\cal{L}}\over \partial A_j}=\partial_i \Big({\partial {\cal{L}}\over \partial \partial_iA_j}\Big),
\end{equation}
yield a Beltrami field; $\nabla\times\mathbf{B}=\lambda \mathbf{B}$. So, in the minimum energy state, the Lorentz force vanishes $\mathbf{j\times B}=0$ which corresponds to the Taylor relaxed state (Taylor 1974). If we write the energy $u_B=\int_V B^2/2$ for a Beltrami field, the coefficient $\lambda$ is found to be proportional to the ratio of magnetic energy to the magnetic helicity, $\lambda=2u_B/|{\cal{J}}_M|$, which determines the force-free configuration. In a diffusive medium, $\eta\neq 0$, the state in which magnetic field is parallel to the current, implied by $\mathbf{j\times B}=0$, translates into a significant dissipation of magnetic helicity according to eq.(\ref{helicity3}). On the other hand, for a given diffusivity, magnetic helicity is conserved when the current helicity vanishes $\mathbf{B.}\nabla\times\mathbf{B}=0$. But with a non-zero Lorentz force, the plasma relaxes to a minimum energy state during which the magnetic helicity will also dissipate, although with a lower rate than the energy (see also \cite{Blackman2015}). In fact, the magnetic energy corresponding to a force free magnetic field in closed systems is a minimum. This is known as Woltjer's theorem \cite{Woltjer1958}.

\subsection{Lagrangian Formalism}

The simple variational treatment of the Taylor relaxation may look motivating for constructing a Lagrangian formalism also for the conservation of magnetic helicity. The evolution equation of magnetic helicity in ideal MHD, $\partial_\mu{J}^\mu_M=0$, is of course very similar to the continuity equation for electric charge, $\partial_\mu \mathbf{j}^\mu=0$ where $\mathbf{j}^\mu=(\rho,\mathbf{j})$ is electric current four-vector. In the language of gauge theories, the gauge invariance of the Lagrangian leads to the conservation of electric charge that corresponds to the so-called $U(1)$ symmetry. The physical implication is that the creation of a net electric charge, at some point of space, could be used to measure the potential which is not allowed. This symmetry appears in the electromagnetic Lagrangian which is constructed by the available scalar Lorentz invariants and an interaction term with matter. Although, there have been attempts to add the Lorentz invariant $\mathbf{E.B}$ to the Lagrangian, for example in magnetogenesis theories, but this would break the parity symmetry unlike the other scalar invariant $B^2-E^2=F^{\mu\nu}F_{\mu\nu}/2$. The electromagnetic Lagrangian reads
\begin{equation}\label{lagrange2}
\begin{split}
{\cal{L}_{EM}}&=-{1\over 4}F_{\mu\nu}F^{\mu\nu}+A_\mu j^\mu\\
&=-{1\over 2}(\partial_\mu A_\nu)^2+{1\over 2}(\partial_\mu A_\mu)^2+A_\mu j^\mu.
\end{split}
\end{equation}

One may fix the gauge in the above Lagrangian before getting the equations of motion. For example, for the Lorentz choice, we either put $\partial_\mu A^\mu=0$ or add an $R_\xi$-gauge term $-(\partial_\mu A^\mu)^2/2\xi$ to the Lagrangian with Feynman-'t Hoof choice $\xi=1$. The field tensor $F_{\mu\nu}$ and the current $\mathbf{j}_\mu$ are invariant under the gauge transformation $A_\mu\rightarrow A_\mu+\partial_\mu \Lambda$. To make the Lagrangian also invariant under this transformation, the additional term in the action, $j^\mu\partial_\mu \Lambda$, must vanish which, after integrating by parts, leads to the continuity equation for the electric charge, $\partial_\mu j^\mu=0$. The analogy of the latter with the helicity equation in ideal MHD, to the helicity equation in ideal MHD, so it might seem suggestive of a similar Lagrangian approach to the latter as well. Since the continuity equation corresponds to the gauge invariance of the Lagrangian through the interaction term $A_\mu j^\mu$, one may ask if there is any symmetry of an MHD Lagrangian that leads to the conservation of magnetic helicity. One might even naively attempt to consider the term ${A}_\mu{J}_M^\mu$ for magnetic helicity as the counterpart of the interaction term ${A}_\mu{j}^\mu$ in electromagnetism. Nevertheless, as a matter of fact, the magnetic helicity is gauge dependent and we would have serious difficulties in obtaining it through a gauge symmetry. In fact, the latter quantity vanishes:

\begin{equation}
{A}_\mu{J}_M^\mu=0,
\end{equation}
which also can be combined with the helicity equation to give us a vanishing "covariant derivative";
\begin{equation}
D_\mu J_M^\mu=(\partial_\mu-ie{A}_\mu){J}_M^\mu=0,
\end{equation}
with $e$ being the electron's charge. However, this does not seem something useful or promising. Another identity is
\begin{equation}
{A}^\mu{B}_\mu+\gamma {J}_M^\nu{U}_\nu=0.
\end{equation}

One also can write a Lagrangian whose corresponding equations of motion (instead of any gauge invariance) yield the magnetic helicity equation, eq.(\ref{helicity-1}). Before doing this, let us consider a subtlety in covariant formalism of electromagnetism. The electromagnetic Lagrangian, given by eq.(\ref{lagrange2}), yields only one pair of the Maxwell's equations. The other pair comes from the definition of the fields in terms of the potentials for the definitions $\mathbf{E}=-\nabla\phi-\partial_t \mathbf{A}$ and $\mathbf{B}=\nabla \times \mathbf{A}$ automatically lead to $\nabla\times\mathbf{E}=-\partial_t \mathbf{B}$ and $\nabla\mathbf{.B}=0$. Thus, there is no need to include them in the Lagrangian. Nevertheless, if we still insist to do so, the "helicity equation" given by (\ref{helicity-1}) is a good choice as a Lagrangian:
\begin{equation}\notag
{\cal{L}}=-2\mathbf{E.B}=2\Big(\nabla\phi+{\partial\mathbf{A}\over\partial t}   \Big).\nabla\times \mathbf{A}.
\end{equation}
It is easy to check that variation with respect to $\phi$ leads to $\nabla.\mathbf{B}=0$ whereas variation with respect to $\mathbf{A}$ gives rise to $\nabla\times\mathbf{E}=-\partial_t \mathbf{B}$. Nevertheless, we emphasize that this Lagrangian is a "redundant" theory as the equations of motion are already set up as definitions. In a similar manner, an MHD Lagrangian that gives rise to the helicity equation (\ref{helicity-1}) would contain redundant pieces since that equation is indeed an identity already built in into the structure of electromagnetic fields. However, in MHD, we invoke a constraint on $\mathbf{E.B}$ which results from the generalized Ohm's law, $\mathbf{E}+\mathbf{v\times B}=\eta\mathbf{j}$ or for the ideal MHD, $\mathbf{E}+\mathbf{v\times B}=0$ (or equivalently $U^\mu B^\nu-U^\nu B^\mu=0$). Therefore, just as a mathematical trick, we may think of $-2\mathbf{E.B}$ as a source term for the magnetic helicity and build a Lagrangian formulation for the helicity equation given by (\ref{helicity-1}). Consider the Lagrangian 
\begin{equation}\label{L_EM}
{\cal{L}}=J_M^\mu\partial_\mu \phi+2 \phi {\partial \mathbf{A}\over\partial t}.\mathbf{B}.
\end{equation}
We have
\begin{equation}
{\cal{L}_{EM}}={\partial \phi\over \partial t} \Big(\mathbf{A}.\mathbf{B}\Big)+\nabla\phi. \Big( \mathbf{A}\times {\partial\mathbf{A}\over\partial t}+\phi \mathbf{B} \Big)+2\phi {\partial \mathbf{A}\over\partial t}.\mathbf{B}.
\end{equation}
The corresponding equation of motion with the dynamic variable $\phi$,
\begin{equation}\notag
{\partial {\cal{L}} \over\partial\phi}={d\over dt}{\partial {\cal{L}}\over \partial \dot \phi}+\nabla.{\partial {\cal{L}}\over \partial \nabla\phi}  ,
\end{equation}
yields
\begin{equation}\notag
-2\mathbf{E.B}={\partial \mathbf{A.B}\over\partial t}+\nabla.\Big(\mathbf{E}\times\mathbf{A}+\mathbf{B}\phi  \Big),
\end{equation}

which is the helicity equation, eq.(\ref{helicity-1}).

\subsection{Spatial Complexity of Magnetic Fields}\label{SpatialComplexity}

We argued that magnetic helicity is a measure of twistedness of magnetic field lines. There is another similar concept, spatial complexity, which is a measure of self-entanglement and complexity of the field. This concept can be generally applied to any vector field. For instance, one can use this quantity to quantify the level of spatial complexity or randomness associated with the fluid motions in a turbulent river. As an application, consider the fact that the complexity of solar magnetic fields increases in a stochastic manner until a sudden change in their topology leads to smoother configurations: how can we quantify this magnetic complexity and randomness and relate them to magnetic topology change? Such an approach has potential applications in a diverse range of problems such as star formation, turbulence and weather forecasting. Previous work \citep*{JV2019} has provided rigorous mathematical definition for magnetic stochasticity level in terms of its renormalized, i.e., coarse-grained, components at different scales. Magnetic field ${\bf B}({\bf x}, t)$ is coarse-grained, or renormalized, at scale $l$ by multiplying it by a test function $G_l({\bf r})$ and integrating: 

$${\bf B}_l({\bf x}, t) =\int_V G_l({\bf r}){\bf B}({\bf x + r}, t)d^3r.$$

Physically, this quantity can be interpreted as the average magnetic field of a parcel of fluid of length-scale $l$ at point $({\bf x}, t)$. Here, $G_l({\bf{r}}) =l^{-3} G({\bf{r}}/l) $ with $G({\bf{r}})$ is a smooth, rapidly decaying kernel. Without loss of generality, we may assume $G({\bf{r}})\geq 0$, $\lim_{|\bf r|\rightarrow \infty} G({\bf{r}})\rightarrow 0$, $\int_V d^3r G({\bf{r}})=1$, $\int_V d^3r \; {\bf{r}}\;G({\bf{r}})=0$, $\int_V d^3r |{\bf{r}}|^2 \;G({\bf{r}})= 1$, and $G({\bf{r}})=G(r)$ with $|{\bf{r}}|=r$, i.e. isotropic kernel, which leads to $\int d^3r\;r_i r_j G({\bf{r}})=\delta_{ij}/3$ \cite{JV2019}. The renormalized field ${\bf{u}}_l$ represents the average field in a parcel of fluid of length scale $l$ at position $\bf x$.

In general, ${\bf B}_l({\bf x}, t)$ will differ from ${\bf B}_L({\bf x}, t)$ for $l\neq L$. For a stochastic field ${\bf B}({\bf x}, t)$ (in turbulence), the angle between ${\bf B}_l({\bf x}, t)$ and ${\bf B}_L({\bf x}, t)$ at any arbitrary point ${\bf x}$ will fluctuate as a stochastic variable. Therefore, $\phi({\bf x}, t) = \cos\theta = \hat{\bf B}_l.\hat{\bf B}_L$ is a measure of local magnetic stochasticity at point $({\bf x},t)$. The rms-average of $(1-\phi)/2$ is a time-dependent, volume-averaged function which measures magnetic stochasticity level in a volume $V$: $S(t) = (1-\phi)_{rms}/2$. The temporal changes in the stochasticity level in turn define topological deformations of the magnetic field and can be related to magnetic topology change. A short review of these concepts is given the following.

The \textit{scale-split energy density}, $\psi({\bf{x}}, t)$, is defined \cite{JV2019} in terms of the renormalized vector field ${\bf{B}}_l({\bf{x}}, t)$ at scale $l$ and the renormalized field ${\bf{B}}_L({\bf{x}}, t)$ at scale $L$ as
\begin{equation}
\psi({\bf{x}}, t)={1\over 2} \;{\bf{B}}_l({\bf{x}}, t){\bf{.B}}_L({\bf{x}}, t).
\end{equation}
which can be divided up into two other functions as $\psi({\bf x}, t)=\phi({\bf x}, t)\chi({\bf x}, t)$ where
 
 \begin{equation}\label{phichi1}
\phi({\bf{x}}, t)=\begin{cases}
\hat{\bf{B}}_l({\bf{x}}, t).\hat{{\bf{B}}}_L({\bf{x}}, t) \;\;\;\;\;\;B_L\neq 0\;\&\;B_l\neq0,\\
0\;\;\;\;\;\;\;\;\;\;\;\;\;\;\;\;\;\;\;\;\;\;\;\;\;\;\;\;\;\;\;\;\;otherwise.
\end{cases}
\end{equation}
which is called the topology field 
and

\begin{equation}\label{phichi2}
\chi ({\bf{x}}, t)={1\over 2} B_l ({\bf{x}}, t) B_L({\bf{x}}, t).
\end{equation}
which is called cross energy field.

The spatial complexity $S_2$, can also be used as stochasticity level for stochastic magnetic fields in turbulent systems;
\begin{equation}\label{formulae}
S_2(t)={1\over 2} (\phi-1  )_{rms}.
\end{equation}

Topological deformation $T_2=\partial_t S_2(t)$, mean cross energy density $E_2(t)$, and field dissipation $D_2=\partial_t E_2(t)$ are given by (for more general definitions see \citep*{JV2019})

\begin{equation}\label{Tdeform}
T_2(t)=  {1\over 4 S_2(t)} \int_V \;(\phi-1){\partial \phi\over \partial t}\; {d^3x\over V},
\end{equation}

\begin{equation}
E_2(t)=\chi_{rms},
\end{equation}

and
\begin{equation}
D_2(t)={1\over  E_2(t)}\int_V \chi \partial_t \chi{d^3x\over V}.
\end{equation}

We will not consider the details of this formalism here. Suffice to say that one can renormalize the induction and Navier-Stokes equations to obtain explicit expressions for the stochasticity and other quantities defined above. It turns out that this formalism provides a potentially powerful way to study magnetic field generation, i.e., dynamo action, magnetic reconnection and other similar problems. A detailed consideration of these concepts can be found in \cite{JV2019}. Application in MHD turbulence and reconnection along with numerical simulations can be found in \cite{SecondJVV2019}. Also a quantitative relationship between magnetic stochasticity and magnetic diffusion has been developed and numerically tested in MHD turbulence by \cite{JVV2019}.

\subsection{Renormalization and Scale Separation}\label{ScaleSeparation}

For a uniformly Lipschitz continuous field ${\bf B}({\bx}, t)$, with renormalized components ${\bf B}_l$ and ${\bf B}_L$ at scales $L>l\neq 0$, consider the following double limit for the corresponding scale split energy density $\psi_{l, L}$:
\begin{equation}\notag
{\psi}_\infty({\bf{x}}, t):=\lim_{l\rightarrow 0} \lim_{L\rightarrow \infty} \psi_{l,L}({\bf{x}}, t)={1\over 2} \;\underbrace{  {\bf{B}}({\bf{x}}, t)}_\text{local\;field}.\underbrace{ \overline{\bf{B}}({\bf{x}}, t) }_\text{global\;field},
\end{equation}
where\footnote{The limiting case of the scale split energy density, 
$$\psi_\infty({\bx}, t)={1\over 2}{\bf B}({\bx}, t).\int_V{\bf B}({\bx-\bx'}, t) {d^3x'\over V},$$
provides a measure for the field's spatial complexity. This scalar field is somehow similar to magnetic helicity density $H=\bf A.B$ which is also a scalar field and measures the field's twistedness and knottedness. In the Lorenz gauge, it is
$$H({\bx}, t)={\bf B}({\bx}, t).\int_V{  \nabla_{\bx'}\times  {\bf B}({\bx'}, t)  \over |{\bx-\bx'}|    } d^3x'.$$
} 
$$\lim_{l\rightarrow 0} {\bf B}_l({\bx}, t)\rightarrow {\bf B}({\bx}, t),$$
and

$$\lim_{L\rightarrow\infty} {\bf B}_L({\bx}, t)\rightarrow \overline{\bf{B}}({\bx}, t)=\int_V  {\bf{B}}({\bx+\br}, t) {d^3 r\over V}$$

is the spatial average of ${\bf{B}}$ at $({\bx}, t)$ defined in an arbitrary volume $V$. The above expression follows from the assumptions that the test function $G(r/l)=l^3 G_l({\bf r})$ is a rapidly decaying, symmetric and smooth test function which is  normalized to unity. Next, we note that in the limit when $L\rightarrow \infty$ and $l\rightarrow 0$, the high-pass filter $\delta {\bf}{\bf B}_{l, L}({\bx}, t)={\bf B}_l({\bx}, t)-{\bf B}_L({\bx}, t)\geq 0$ defines a fluctuating component;

$$\delta {\bf B}({\bx}, t)={\bf B}({\bx}, t)-\overline{\bf B}({\bx}, t).$$
with a vanishing average, $\overline {\delta {\bf B}}=0$. Therefore, coarse-graining, in the limit $l\rightarrow 0$ and $L\rightarrow \infty$, naturally leads to a simple scale separation scheme;
$${\bf B}=\overline{\bf B}+\delta{\bf B},\;\;\;\overline{\delta{\bf B}}=0.$$

In some applications, e.g., in cosmology, it may be convenient to use a different decomposition, which highlights the role of isotropy and homogeneity, by writing the arbitrary field ${\bf B}({\bf x}, t)$ in terms of a homogeneous ${\bf B}(t)$ and an inhomogeneous ${\bf B}'({\bf x}, t)$ part; 

$${\bf B}({\bf x}, t)={\bf B}(t)+{\bf B}'({\bf x}, t),$$

 where 
 
 $${\bf B}(t)=\int_V {\bf B}({\bf x}, t) {d^3x\over V}$$ 
 
 and 
 
 $$\int_V {\bf B}'({\bf x}, t) {d^3x\over V}=0.$$
 In general, for an isotropic field, ${\bf B}(t)=0$, while for a homogeneous field, ${\bf B}'({\bf x}, t)=0$. Such decompositions are widely used, e.g., in cosmological models in which the universe is assumed to be homogeneous and isotropic at large scales.
It also follows that the renormalized field ${\bf B}_l({\bf x}, t)$ is consequently decomposed into a homogeneous, ${\bf B}_l(t)$, and an inhomogeneous part ${\bf B}'_l({\bf x}, t)$;

\begin{equation}\notag
{\bf B}_l({\bf x}, t)={\bf B}_l(t)+{\bf B}'_l({\bf x}, t),
\end{equation}
where we have
\begin{equation}\notag
{\bf B}_l(t)=\int_V  {\bf B}_l({\bf x}, t) {d^3x\over V},
\end{equation}
and
\begin{equation}\notag
\int_V  {\bf B}_l'({\bf x}, t) {d^3x\over V}=0.
\end{equation}

It is common to appeal to scale separation in order to take the role of turbulence into account. However, as already discussed above, there are more precise mathematical methodologies, e.g., in terms of renormalization \cite{JV2019}. Here, we will briefly discuss the application of conventional scale separation to magnetic helicity. 

One can define the large scale and small scale magnetic helicities respectively as $\overline{J^0_M}= \overline{\mathbf{A}}.\overline{\mathbf{B}}$ and $\delta{J}^0_M=\overline{ \delta{\mathbf{A}}.\delta{\mathbf{B}}} $.  Note that since the helicity is quadratic in fields, its separation into turbulent and mean parts must be used with care as we have $\overline{ \delta{J}^0_M}\neq 0$ and in fact $\overline{ \delta{J}^0_M }=\delta{J}^0_M$ is an averaged quantity. Applying this to eq.(\ref{helicity3}), we arrive at

\begin{equation}\label{helicity5}
{\partial \overline{J^0_M}\over\partial t}+\nabla.\overline{\mathbf{J}}_M=2{\cal{E}}.\overline{\mathbf{B}}-2\eta\overline{\mathbf{j}}.\overline{\mathbf{B}},
\end{equation}
and 
\begin{equation}\label{helicity6}
{\partial \delta{J^0_M}\over\partial t}+\nabla.\delta{\mathbf{J}}_M=-2{\cal{E}}.\overline{\mathbf{B}}-2\eta\overline{\delta{\mathbf{j}}.\delta{\mathbf{B}}},
\end{equation}
where we have introduced the electromotive force ${\cal{E}}=\overline{ \delta\mathbf{v}\times\delta\mathbf{B}}$ and also the mean and turbulent helicity flux terms:
\begin{equation}\label{helicity7}
\overline{\mathbf{J}}_M= \overline{\mathbf{A}}\times\Big(\overline{\mathbf{v}} \times \overline{\mathbf{B}}+{\cal{E}}-\eta \overline{\mathbf{j}} +\nabla\overline \phi \Big) ,
\end{equation}
and
\begin{equation}\label{helicity8}
\delta{\mathbf{J}}_M=\overline{ \delta{\mathbf{A}}\times\Big(\overline{\mathbf{v}} \times \delta{\mathbf{B}}+\delta{\mathbf{v}} \times \overline{\mathbf{B}}-\eta \delta{\mathbf{j}} +\nabla\delta \phi \Big) }.
\end{equation}
In the ideal MHD, one works in the limit of vanishing resistivity $\eta\rightarrow 0$ assuming that the equations remain well-defined (e.g., in the presence of turbulence, taking this limit will lead to singular fields which makes MHD equations ill-defined; see \cite{JV2019} and \cite{Eyink2018} and references therein). Note that in the above expressions, we have ignored the terms with vanishing divergence as they play no role in our considerations. It is already obvious from these expressions, that the magnetic helicity is deeply connected to the electromotive force, ${\cal{E}}$, which is the most important quantity in dynamo theories to be discussed presently.

\section{Helicity in Dynamo Theories}
A magnetic dynamo mechanism seems to be at work in almost all types of celestial bodies such as stars, galaxies and accretion discs. A dynamo generating a large scale field needs an electrically conducting fluid, a source of kinetic energy, and an internal energy source. Thermal energy can be generated by several sources such as gravitational potential and radioactive reactions.  Kinetic energy too can be provided by, for example, rotation of the celestial body. In rotating celestial bodies like stars and planets this type of energy is easily available of course. The internal source of energy is a requirement in order to move the conducting fluid which provides convection. Yet, there is another important ingredient to dynamos: seed fields. A dynamo action requires, among other aforementioned constituents, also a seed magnetic field to begin with. The initial seed field, which is typically weak compared with the dynamo generated field, can be an external one penetrating the system or a fossil field which has been sitting there since the formation of the system. Indeed, the fossil fields were, and to a lesser degree are, one alternative hypothesis in explaining the large scale magnetic fields of astrophysical objects. This idea leads us to at least two other questions. How was the fossil field itself generated in the first place and how did it resist dissipation in such long times? To answer the former, one eventually has to go back to the early universe and study magnetogenesis which is not part of our consideration here. The latter, dissipation over long time scales, is a substantial objection to this alternative to dynamo action. At most, they would serve as weak seed fields for dynamos that kept generating the stronger large scale fields (scales comparable to the size of the system). This dynamo mechanism, operating in the variety of celestial bodies, magnetizes our universe over a wide spectrum of length scales.
\iffalse
 \begin{figure}
\includegraphics[scale=.5]{Pictures/BTfields.jpg}
\centering
\caption {\footnotesize {An illustration of an $\alpha\Omega$ dynamo. An initial poloidal magnetic field threads a rotating body of highly conductive fluid (a). Differential rotation and shear stretch and bend the field lines generating a toroidal component (b,c, and d). Turbulence then can take action and re-generate a poloidal field through the $\alpha$-effect (e) to complete the dynamo cycle (f). Illustration from Love (1999).}}
\label{BTfield}
\end{figure} 
\fi
Given, a large (small) scale dynamo action can generate large (small) scale magnetic fields but where does magnetic helicity come into play in this theory? The answer is many places. To set an example and get an insight, consider a rotating body of hot and electrically conductive plasma like a star. Obviously, solid body rotation is irrelevant for a hot body of conductive fluid. A more realistic situation would include differential rotation which also gives rise to shear between stellar layers. High electrical conductivity means magnetic field lines are almost "frozen in" into the plasma. Thus shear and differential rotation stretch and bend  the seed poloidal field and generate a toroidal field. This is half a dynamo cycle though: a successful dynamo action must make this toroidal field generate a poloidal component in turn to complete the cycle. Magnetic buoyancy, for example, can rise the toroidal flux tubes toward the surface. On their way to the surface, turbulence takes the scene and affects the flux tubes: helical motions can bend and twist the field lines to create an alpha shape (alpha effect). This gives rise to a poloidal field and finishes the cycle. Such an $\alpha\Omega$ dynamo seems promising and practical but we have to pause here to ponder the helicity of the system. We had to stretch and twist the toroidal field in order to generate the poloidal component which means we have created magnetic helicity in large scales. This looks alright but not quite. Magnetic helicity is conserved, so how could we generate some excess amount of it? Short answer is that, roughly speaking, a small scale magnetic helicity has in fact been also generated with opposite sign which compensates the newborn large scale helicity. Magnetic field lines may be imagined then as the filament of a light bulb which is in fact multiple coils of a coiled fine wire. Large (small) scale twists correspond to large (small) scale magnetic helicity. The bottom line is that dynamo requires to generate large scale magnetic helicity in converting toroidal-poloidal field components. However, conservation of magnetic helicity becomes a problem for, it turns out, the small scale magnetic helicity destroys the dynamo at large magnetic Reynolds numbers. This quenching effect is another reason why magnetic helicity is believed to play a critical role in dynamo theories. 

Gruzinov and Diamond \cite{Gruzinovetal1994} showed that the accumulation of small scale magnetic helicity would quench kinematic dynamos (briefly discussed in \S2) and obtained an expression for alpha quenching. This work may be regarded as a starting point considering that, since then, there has been a great deal of work aimed at incorporating magnetic helicity into dynamo theories, especially in the context of dynamic alpha quenching. One line of work has been focused around possible ways of removal of the unwanted small scale magnetic helicity to alleviate the quenching process. This helicity flux expulsion may be a naive, or at least oversimplified, part of the picture since it is not clear how the system chooses the unwanted part to eject. Moreover, Vishniac and Cho \cite{VishniacandCho2001} showed that a dynamo can work in a periodic box with no helicity flux expulsion at all. Furthermore, the ejected helicity, if any, may be contained in both large and small scale structures so the net magnetic helicity can remain unchanged. However, since ejecting magnetic helicity bound up in large scale structures would require much less energy, it may be the preferred mode of ejection \cite{Vishniac2009}. In any case, the observed fast generation of large scale fields may depend on the generation of large scale helicity flux from eddy scale processes \cite{VishniacandCho2001}.

William Gilbert was probably the first person who proposed a scientific explanation for the origin of our planet's magnetic field: natural permanent magnetism. However, the Earth's core is too hot for the permanent magnetism to survive. Einstein suggested that a hypothetical asymmetry between the charges of electrons and protons, partly making up the rotating Earth, could produce the observed magnetic field. Developments in particle physics, of course, rejected this idea. Later on, in 1940's, Walter Els\"{a}sser proposed that the conducting fluids in the Earth's outer core could be responsible for generating the magnetic field. Some necessary tools were in fact developed decades earlier by Joseph Larmor who had already put forward a dynamo mechanism to explain the sunspots in 1919. The extensive studies followed after the works of these two pioneers, extended also to explain magnetic fields of other celestial bodies, have collectively become known as the magnetic dynamo theories.

The simplest configuration capable of generating a large scale dipole-like magnetic field requires a helical motion of electrically conducting material. So, one might naively think of a circulating current in a fluid as a potential magnetic dynamo. But a self-sustaining dynamo must also survive the resistive dissipation which makes the required velocity field much more complicated. Thus, it is tempting to think of a velocity field $\mathbf{v}$ in a conductive fluid that can sustain a dynamo action by converting kinetic energy to magnetic energy. One difficulty arises, however, from a back-reaction effect since the generated magnetic field would not remain passive. Starting with a fixed given velocity field, the generated magnetic field would in turn interact with it through the Lorentz force. At later times, when the field becomes strong enough, this changes the initial velocity field making it a non-linear problem. Turbulence, ubiquitous in astrophysics, will also affect the velocity field in a significant way. In fact, even if conceivable, such a "kinematic dynamo" would die young long before magnetic field reaches equipartition with the velocity field. Another approach is to let the magnetic and velocity fields evolve and interact in a turbulent medium and look for exponentially growing solutions of the magnetic field. This is the main idea behinf mean field dynamo theories discussed presently.

\subsection{Mean Field Theory}

A real magnetic field configuration, e.g., in stars and accretion disks, will be extremely complicated as the differential rotation, convection and different instabilities become involved affecting the velocity and magnetic fields. Nevertheless, one may argue that a "statistical" approach based on the scale separation in a turbulent medium might still reflect the basic underlying physics. Scale separation (see \S\ref{ScaleSeparation}), however, can be replaced with mathematically more concise methodologies e.g., coarse-graining, briefly discussed in \S\ref{SpatialComplexity} (see e.g., \cite{Eyink2015}; \cite{JV2019}).  In the following, we will review the conventional mean field theory assuming scale separation.

The velocity field, governed by the Navier-Stokes equation, couples with the magnetic field also through the induction equation. After dividing the fields into the mean and turbulent parts, as before, we find 
\begin{equation}\label{induction1}
{\partial\overline{\mathbf{B}}\over\partial t}=\eta \nabla^2\overline{\mathbf{B}}+\nabla\times(\overline{\mathbf{v}}\times \overline{\mathbf{B}})+\nabla\times\cal{E},
\end{equation}
with ${\cal{E}}=\langle \delta\mathbf{v\times \delta B}\rangle$ and
\begin{equation}\label{induction2}
{\partial\delta\mathbf{B}\over\partial t}=\eta \nabla^2\delta\mathbf{B}+\nabla\times(\delta{\mathbf{v}}\times \overline{\mathbf{B}}+\delta\mathbf{v}\times\overline{\mathbf{B}}).
\end{equation}
The term $\langle\delta\mathbf{v}\times\overline{\mathbf{B}}\rangle$ indicates that the
mean magnetic field contributes to the generation of the turbulent field $\delta\mathbf{B}$. Consequently, an initial condition such as $\delta\mathbf{B}(t=0)=0$  along with the linearity of eq.(\ref{induction2}) imply a linear relationship between $\overline{\mathbf{B}}$ and $\delta\mathbf{B}$. Therefore, the electromotive force ${\cal{E}}=\langle \delta\mathbf{v\times \delta B}\rangle$ linearly depends on $\overline{\mathbf{B}}$. This suggests, in turn, that we can Taylor expand ${\cal{E}}$ as
\begin{equation}\label{induction3}
 {\cal{E}}_i=\alpha_{ij}\overline B_{j}+\beta_{ijk}\frac{\partial \overline B_{j}}{\partial x_k}+\gamma_{ijkl}\frac{\partial^2 \overline B_{j}}{\partial x_k \partial x_l}+...\; .
 \end{equation}
Keeping only the first two terms, one may calculate the tensor coefficients $\alpha$ and $\beta$. If we assume that velocities are correlated only when they are in the same direction, and also the energy is distributed roughly isotropically among the different modes, then we have;
\begin{equation}\label{4.7+2}
\begin{cases}
{\cal{E}}_i=\alpha_{ij}\overline B_j-\epsilon_{ijk} D \frac{\partial \overline B_k}{\partial x_j},\\
D=\frac{\tau}{3} \overline{ \delta v^2 },\\
\alpha_{ij}=\epsilon_{ilm}\overline{ \delta v_l \frac{\partial v_m}{\partial x_j} } \tau.
\end{cases}
\end{equation}
We can call $\alpha_{ij}$, without the turbulent eddy correlation time $\tau$, the fluid helicity tensor whose trace is the kinetic helicity. For isotropic and reflectionally not symmetric turbulence, we can consider the coefficients as scalars and write
\begin{equation}\label{4.5}
{\cal{E}}=\alpha \overline{\mathbf{B}}-\beta \overline{\mathbf{j}},
\end{equation}
where the scalar coefficient $\alpha$ is proportional to the kinetic helicity whereas $\beta$, also known as the turbulent diffusivity, is proportional to the turbulent energy density:

 \begin{figure}
\includegraphics[scale=.45]{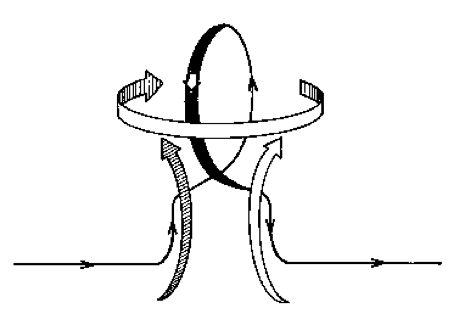}
\centering
\caption {\footnotesize {An illustration of the so-called $\alpha$-effect. In a turbulent flow, a typical magnetic field line can rise and then become twisted and bent creating an $\alpha$-like shape. This effect generates a poloidal component out of a toroidal field component. Illustration from \cite{Parker1970}.}}
\label{Alpha-effect.jpg}
\end{figure} 

 \begin{equation}\label{4.6}
\alpha=-\frac{\tau}{3} \overline{ \delta \textbf{v.}\delta {\bf{w}}  },
\end{equation}

\begin{equation}\label{4.7}
\beta=\frac{\tau}{3} \overline{ |\delta \textbf{v}|^2 }.
\end{equation}

One physical interpretation of eq.(\ref{4.5}) is that the so-called $\alpha$-effect is capable of amplifying the seed magnetic field while the $\beta$-effect can enhance the diffusion rate. The back-reaction of the magnetic field modifies the $\alpha$ coefficient by adding a term proportional to current helicity in isotropic turbulence \cite{Pouquetetal1975}:
 \begin{equation}\label{4.7.2}
\alpha=-{\tau\over 3} \Big(  \overline{ \delta \mathbf{v.}\delta {\mathbf{w}}  }- {1\over\rho} \overline{ \delta \mathbf{j.} \delta \mathbf{B} } \Big),
\end{equation}
where $\rho$ is the density.

Thus far, we have not yet considered magnetic helicity in our treatment of the dynamo action. In order to avoid some major discrepancies in the theory, it turns out that we have to take the helicity conservation into our consideration. Magnetic helicity does not directly appear in the above dynamo equations, however, the $\alpha$-effect given by eq.(\ref{4.7.2}), contains the current helicity which is closely related to the magnetic helicity. First of all, we express the current helicity in terms of the vector potential;\begin{equation}\notag
\overline{ \delta \mathbf{j.} \delta \mathbf{B} }=\overline{ \nabla\times\nabla\times\delta\mathbf{A.B}}=-\overline{ (\nabla^2 \mathbf{A}).\mathbf{B}},
\end{equation}
where we have used the Coulomb gauge for simplicity. This quantity is not conserved but we can estimate it with the conserved magnetic helicity. Consider a Fourier transform for the vector potential;
\begin{equation}\notag
\mathbf{A}(\vec{x})=\int \mathbf{A}(\vec{k})e^{i \vec{k}.\vec{x}} d^3k.
\end{equation}
Thus, $\nabla^2\mathbf{A}(\vec{x})\rightarrow k^2 \mathbf{A}(\vec{k})$ using which we find $\overline{ \delta \mathbf{j.} \delta \mathbf{B} }\rightarrow k^2 \overline{J^0_M}$.

We can re-write the evolution equations for the small and large scale magnetic helicities using the expression we found for the electromotive force ${\cal{E}}$:

\begin{equation}\label{helicity25}
{\partial \overline{J^0_M}\over\partial t}+\nabla.\overline{\mathbf{J}}_M=2\alpha\overline{B}^2-2(\eta+\beta)\overline{\mathbf{j}}.\overline{\mathbf{B}},
\end{equation}
and 
\begin{equation}\label{helicity26}
{\partial \delta{J^0_M}\over\partial t}+\nabla.\delta{\mathbf{J}}_M=-2\alpha\overline{B}^2-2(\eta+\beta)\overline{\delta{\mathbf{j}}.\delta{\mathbf{B}}},
\end{equation}
where $\eta_T=\eta+\beta$ is sometimes called the total diffusivity. Note that with a negligible total diffusivity and in a steady state when $\partial_t \delta J_M^0 \sim 0$, we find
\begin{equation}
\alpha\sim -{\nabla.\delta \mathbf{J}_M\over 2{\overline B}^2}.
\end{equation}
However, in a magnetic dynamo, the time evolution of the small scale helicity $\partial_t \delta J_M^0$ in general does not vanish and instead fluctuates. Thus, the above estimate for the $\alpha$-effect is a rough approximation even in the steady state.

\iffalse
 \begin{figure}
\includegraphics[scale=.4]{Pictures/Sun-helicity}
\centering
\caption {\footnotesize {A schematic of the magnetic helicity injection by stellar differential rotation. Illustration from Berger (1999).}}
\label{Sun-helicity}
\end{figure} 
\fi

\begin{figure}
\includegraphics[scale=.5]{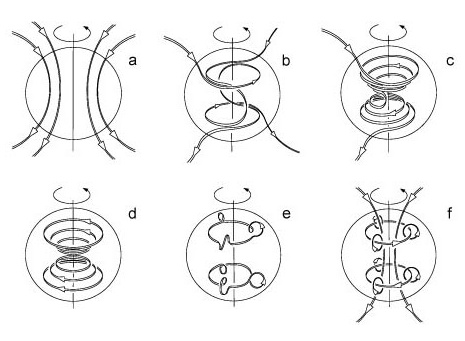}
\centering
\caption {\footnotesize {Stretching of magnetic field lines around a rotating object. Rotation can produce a toroidal field from an initial poloidal field. Turbulence is assumed to be able to react on the produced toroidal field to turn it into a poloidal field. Another step will complete the cycle. Illustration from \cite{Love1999}.}}
\label{Stretching.jpg}
\end{figure}

\subsection{Vishniac-Cho Flux}

Let us consider, as an example, the Vishniac-Cho flux \cite{VishniacandCho2001} in systems with nonhelical turbulence. This flux has an important role in $\alpha^2$ dynamos while its effect on $\alpha\Omega$-dynamos seems negligible. We will use the Coulomb gauge for the vector potential, ${\nabla.}\textbf{A}=0$, for a system with anisotropic turbulence. To use a notation similar to the original literature, here we denote the ensemble averaging as $\overline{X}\equiv \langle X\rangle\equiv X_0$. The electric field $\textbf{E}$ reads
\begin{eqnarray}\notag
\textbf{E}&=&-{\nabla} \Phi -\partial_t \textbf{A}\\ \notag
&=&-(\textbf{v}_0+\delta \textbf{v})\times (\textbf{B}_0+\delta\textbf{B})+\eta(\textbf{j}_0+\delta\textbf{j}) \notag
\end{eqnarray}
 which, taking the average, gives us
  \begin{equation}\label{4.24}
 \textbf{E}_0=-\textbf{v}_0\times\textbf{B}_0- \langle \delta \textbf{v} \times \delta \textbf{B} \rangle+\eta\textbf{j}_0=-{\nabla}\Phi_0-\frac{\partial \textbf{A}_0}{\partial t},
 \end{equation}
 \begin{eqnarray}\notag
 -\delta \textbf{E}&=&\delta \textbf{v} \times \textbf{B}_0+   \textbf{v}_0 \times \delta \textbf{B}  \\\notag
 &\;&+ \underbrace{ \delta \textbf{v} \times \delta \textbf{B}   -\langle \delta \textbf{v} \times \delta \textbf{B} \rangle}_\text{=\textbf{G} }+\eta\delta\textbf{j}\\\label{after4.24}
 &=&-{\nabla}\delta\Phi-\frac{\partial \delta \textbf{A}}{\partial t},
 \end{eqnarray}
where $ \delta \textbf{E}\equiv -\textbf{e}+\eta\delta\textbf{j}$ is the small scale electric field. It is easy to see that in the first order smoothing (i.e. assuming $\delta\bf{v}\times\delta\bf{B}=\langle\delta\bf{v}\times\delta\bf{B}\rangle$) we have $\textbf{e}=\delta \textbf{v} \times \textbf{B}_0+\textbf{v}_0\times \delta \textbf{B}+\textbf{G}\simeq\delta \textbf{v} \times \textbf{B}_0+\textbf{v}_0\times \delta \textbf{B}$, and ignoring the term $\textbf{v}_0\times\delta\textbf{B}_0$ we get $\textbf{e}=\delta \textbf{v} \times \textbf{B}_0=-{\nabla}\delta\Phi-\frac{\partial \delta \textbf{A}}{\partial t}$ with potential $\nabla^2\delta\Phi={\nabla.}(\delta \textbf{v} \times \textbf{B}_0)$. Also, note that with this notation one can take the average of the induction equation to obtain
\begin{equation}\label{4.24+1}
\frac{\partial \textbf{B}_0}{\partial t}={\nabla} \times {\cal{E}}.
\end{equation}
The above equation is the simplest known equation that ${\cal{E}}$ must satisfy. The most common approach in dynamo theory is to make some general assumptions about the structure of turbulence in order to calculate ${\cal{E}}$ as we showed earlier, for example in eq.(\ref{4.5}). However, most of these approaches have a common difficulty; they do not conserve magnetic helicity. Here, our aim is to explore a model of mean field dynamo theory which incorporates the conservation of magnetic helicity following Vishniac and Cho \cite{VishniacandCho2001}. If we assume that the eddy size is much smaller than the scale of the mean magnetic field $\textbf{B}_0$, then we will have
  \begin{equation} \label{4.26}
 \langle \textbf{A.} \textbf{B} \rangle=\langle \textbf{A}_0 \textbf{.B}_0 \rangle,
 \end{equation}
  which is equivalent to assume that the small scale helicity is zero, ${\delta \cal{H}}=\langle \delta \textbf{A.}\delta \textbf{B}\rangle=0$. This assumption holds as long as we deal with an eddy scale smaller than the typical field scale by a factor of at least $(\delta {B}/B_0)^2$ and also there is an efficient transfer of magnetic helicity between scales. Note that because of this assumption, we ignore the term $\delta\textbf{A}\times(\textbf{v}_0\times\delta\textbf{B})$ in the magnetic helicity flux. Under these conditions, the expression (\ref{4.26}) can be justified and we get
\begin{equation}\label{vishniac-cho-1}
2{\cal{E}} \textbf{.B}_0={\nabla.}\langle (\textbf{e}+{\nabla} \delta \Phi) \times \delta \textbf{A} \rangle.
\end{equation}
So
\begin{equation} \notag
{\nabla.}{\delta \textbf{J}}_M\simeq -2{\cal{E}.}\textbf{B}_0.
\end{equation}
We write
\begin{eqnarray} \notag
\delta \textbf{J}_M&=& \langle \delta \textbf{B}\delta \Phi \rangle+\langle \delta \textbf{A}\times[ \textbf{v}_0\times \delta \textbf{B} + \delta \textbf{v}\times\textbf{B}_0] \rangle\\ \notag 
&\equiv&\langle \delta \textbf{A} \times {\nabla} \delta \Phi \rangle+\langle \delta \textbf{A}\times[ \textbf{v}_0\times \delta \textbf{B} + \delta \textbf{v}\times\textbf{B}_0] \rangle\\ \notag
&=&\langle\delta\textbf{A}\times(\textbf{e}+{\nabla}\delta \Phi) \rangle,
\end{eqnarray}
where $\textbf{e}=\textbf{v}_0\times\delta\textbf{B}+\delta \textbf{v} \times \textbf{B}_0$ which yields $\nabla^2\delta\Phi={\nabla.}(\delta \textbf{v} \times \textbf{B}_0)$. We have ignored the term $\delta\textbf{A}\times(\delta\textbf{v}\times\delta\textbf{B})$ since its contribution is a function of only local properties of turbulence (this term is dominant in stratified rotating objects). We can use eq.({\ref{vishniac-cho-1}) to write 
\begin{eqnarray}\notag
{\cal{E}}&=&{\cal{E}}_{\perp}+\frac{\textbf{B}_0}{2B^2_0} {\nabla.} \langle (\textbf{e}+{\nabla} \delta \Phi) \times \delta \textbf{A} \rangle\\ \label{4.29}
&=&{\cal{E}}_{\perp}-\frac{\textbf{B}_0}{B^2_0} {\nabla.} {\cal{J}}_{VC}, 
\end{eqnarray}
where we have written the small scale magnetic helicity flux as 
 \begin{equation}\label{4.29+1}
{\cal{J}}_{VC}=-\frac{1}{2}\langle (\textbf{e}+{\nabla} \delta \Phi) \times \delta \textbf{A} \rangle.
 \end{equation}
In order to evaluate the helicity current term, we can take
\begin{equation}\label{4.30}
\frac{\partial \delta \textbf{A}}{\partial t} \simeq \frac{\delta \textbf{A}}{\tau_c} \simeq (\textbf{e}-{\nabla} \delta \Phi),
\end{equation}
where $\tau_c$ is the eddy correlation time. Then, eq.(\ref{4.29}) can be re-written in terms of the anomalous magnetic helicity flux;
\begin{eqnarray}\notag
{\cal{E}}&=&{\cal{E}}_{\perp}+\frac{\textbf{B}_0}{2B^2_0} {\nabla.} \langle (\textbf{e}+{\nabla} \delta \Phi) \times \delta \textbf{A} \rangle\\ \notag
&=&{\cal{E}}_{\perp}+\frac{\textbf{B}_0}{B^2_0} {\nabla.} \langle {\nabla} \delta \Phi \times \textbf{e} \rangle \tau_c \\ \label{4.31}
&=&{\cal{E}}_{\perp}-\frac{\textbf{B}_0}{B^2_0} {\nabla.} {\cal{J}}_{VC},
\end{eqnarray}
  where we have written the anomalous magnetic helicity current as ${\cal{J}}_{VC}=-\langle    {\nabla} \delta \Phi \times \textbf{e} \rangle \tau_c$. Note that the ansatz (\ref{4.26}) has an important role in the line of reasoning which led to (\ref{4.31}). From eq.(\ref{4.30}) we have ${\nabla.} \textbf{e}=\nabla^2 \delta \Phi$ which can be used in calculating ${\cal{J}}_{VC}$. Using this method, Vishniac and Cho \cite{VishniacandCho2001} evaluated $ {\cal{J}}_{VC}=-\langle  {\nabla} \delta \Phi \times \textbf{e} \rangle \tau_c$ by using a Fourier transformation for $ \textbf{e}$, and then writing $\delta \Phi$ in terms of this Fourier transform. The result is known as the Vishniac-Cho flux:
\begin{equation}\label{4.31+1}
{\cal{J}}_{VC}=-\lambda^2_c \langle (\textbf{B}_0 \textbf{.}\delta{\omega})(\textbf{B}_0 {.\nabla}) \delta \textbf{v} \rangle \tau_c,
\end{equation}
where $\lambda_c$ is an angle-averaged eddy correlation length, and $\delta {\omega}$ is the vorticity. Interestingly, there is no term related to the helicity in the above expression. Nevertheless, there is a physical intepretation behind this expression: the anomalous magnetic helicity current, ${\cal{J}}_{VC}$, is proportional to the correlated product of the gradient of the velocity field along the magnetic field lines, $(\textbf{B}_0 {.\nabla}) \delta \textbf{v}$. This can be interpreted as a fold in the so-called twist and fold theory of Vainshtein and Zeldovich \cite{Vainshteinetal1972}. Also, ${\cal{J}}_H$ is proportional to the vorticity along the magnetic field lines, $ (\textbf{B}_0 \textbf{.}\delta{\omega})$, which can be interpreted as a rotation or twist in the context of the above mentioned theory. It is conventionally assumed that this procedure can repeat itself and create a complete twist and fold dynamo cycle.

Since the build-up of small-scale magnetic helicity leads to the $\alpha$ quenching, so one way for a growing dynamo to survive is the exportation of the magnetic helicity. Observations show that the astrophysical dynamos operating in stars and galaxies and other cosmic objects are not resistively slow (even the transport of external magnetic fields in objects such as accretion disks can be fast, see e.g., \cite{JafariVishniac2018disks}; \cite{Jafari2019} and references therein) . These systems saturate and evolve on dynamical time scales. This may be achieved by expelling magnetic helicity from the domain through helicity fluxes \cite{Brandenburg2009}. We can look at the saturation process as caused by the buildup of magnetic helicity. Then, it would be possible to increase the rate at which large scale field saturates. It is conceivable to reach significant saturation field strengths through mechanisms that export or destroy magnetic helicity \cite{HubbardandBrandenburg2011}. The fast generation of large scale magnetic fields, observed in astrophysical objects, depends on the generation of large scale helicity flux from eddy scale processes \cite{VishniacandCho2001}. However, the ejected magnetic helicity can be contained in both large and small scale structures so net magnetic helicity can remain unchanged. Since ejecting magnetic helicity bound up in large scale structures needs much less energy, it may be the preferred mode of ejection \cite{Vishniac2009}.

The $\alpha$-effect generates a helical large scale magnetic field which means the presence of a large scale magnetic helicity, $\textbf{A}_0\textbf{.}\textbf{B}_0$. On the other hand, since the total magnetic helicity $\textbf{A}_0\textbf{.}\textbf{B}_0+\delta\textbf{A}\textbf{.}\delta\textbf{B}$ is conserved, so a small scale magnetic helicity of opposite sign, $-\delta\textbf{A}\textbf{.}\delta\textbf{B}$, must be generated. The small scale magnetic helicity is responsible for the creation of a magnetic $\alpha$-effect, often represented by $\alpha_M$. Therefore, the magnetic helicity may accumulate locally at small scales. This can lead to quenching of dynamo action at  larger scales of magnetic field which is called $\alpha$ quenching. Indeed, using this idea, Kleeorin and Ruzmaikin (see \cite{Kleeorinetal2000} and references therein) proposed the dynamical quenching even before any numerical simulations indicated  its existence. For high magnetic Reynolds numbers, $\alpha_M$ can get very large which in turn decreases the efficiency of the dynamo action. In real astrophysical bodies like stars, have high magnetic Reynolds numbers with an efficient dynamo action. One possible solution is exporting the small scale magnetic helicity out of the system. The accumulated magnetic helicity is transported either to the regions with helicity of the opposite sign or beyond the borders of the system. 

There are different possible ways, in astrophysical bodies, for removing magnetic helicity from the system. Examples include the turbulent-diffusive magnetic helicity flux and advective flows or winds. In the kinematic theory, the growth of the large scale magnetic field depends on the volume average of the kinetic helicity. In the MFT, we need to inject kinetic helicity at sufficient amplitudes, in the turbulence, to overcome turbulent dissipation. In this case, if the forcing is non-helical, we would not be able to get a large scale magnetic field. Nonetheless, even if we do not inject  kinetic helicity externally, it can still be generated through the forces produced by the small scale current helicity. This is because the small scale current helicity is nonzero in the presence of magnetic helicity. In addition, it is safe to say that, current helicity itself can drive a dynamo. As Vishniac and Cho \cite{VishniacandCho2001} pointed out, the induced transport of magnetic helicity may generate a dynamo even when the kinetic helicity is very small. Consequently, the only requirements for a large scale dynamo are large scale shear and an anisotropic distribution of turbulent power in Fourier space. Thus, utilizing the concept of dynamical $\alpha$-quenching in real dynamos requires a deep understanding of magnetic helicity fluxes, especially that of small scale magnetic helicity.

\subsection{Dynamo Action and Rotation}

An empirical relationship between X-ray luminosity and the Rossby number has been suggested for low-mass stars (see e.g., \cite{Pallavicinietal1981}; \cite{Reinersetal2014}). For low and moderate mass stars, the X-ray luminosity as indicator of magnetic flux increases with rotation until it saturates and does not grow anymore (\cite{Pizzolatoetal2003}; \cite{Christensenetal2009}). A successful dynamo theory is thus expected to connect the large scale stellar magnetic field to x-ray activity, rotation and shear. Galaxies and accretion discs should be included as well and the dynamo activity of these rotating objects should somehow be related to the rotation. Rossby number is a useful parameter in quantifying such relationships.

It turns out that the required electric fields, for generation of large scale magnetic fields in dynamos, have a deep connection with the transfer of magnetic helicity between scales. This fact has led to attempts to understand dynamo theory in terms of magnetic helicity---a concept that quantifies the way magnetic field wraps around itself. Since the energy corresponding to a given magnetic helicity scales inversely with the scale of the twists, the magnetic helicity has a tendency to accumulate at the largest scales (see e.g., \cite{ShapovalovandVishniac2011}). In other terms, magnetic helicity has an inverse cascade \cite{Frischetal1975}. Large scale magnetic fields, generated via dynamo action, must have both poloidal and toroidal components in order to compensate diffusive losses. Hence, the system needs a non-zero magnetic helicity. The translation of $\alpha$-quenching in the language of magnetic helicity is that the $\alpha$-effect generates magnetic helicity in large scales. Because of the conservation of magnetic helicity, however, the dynamo must also generate a small scale magnetic helicity with the opposite sign to keep the total helicity conserved. This small scale component can eventually become so large and capable of quenching the dynamo action at large magnetic Reynolds numbers.

A satisfactory dynamo theory must be more than internally self-consistent. It should make predictions and give insight into observations of magnetic fields in stars, galaxies and other cosmological objects. For instance, at small Rossby numbers or roughly speaking the fast rotation regime, the stellar activity indicators saturate and do not grow over a certain level. Underlying reason may be the saturation of the dynamo action. Or else the dynamo may continue to operate but the stellar surface may be totally covered by spots leaving no space for emitting plasma with excess field. We need to establish relationships between magnetic field, density, rotation, shear and a typical length scale of the system \cite{Reinersetal2014}. As an example of this type of correlations, Frick et al. \cite{Fricketal2016} have recently shown that the spatial power spectra of several spiral galaxies, which contain magnetic arms, are similar for the tracers of dust, gas, and total magnetic field. Nevertheless, the spectra of the ordered magnetic fields in these galaxies are found to be very different. In another recent paper, Blackman and Thomas \cite{BlackmanandThomas2015} have related the stellar coronal activity and Rossby number in the late type main sequence stars by constructing an expression for coronal luminosity based on dynamo field generation and magnetic buoyancy."

Blackman and Thomas \cite{BlackmanandThomas2015} explained the observed dependence of X-ray/bolometric luminosities on the Rossby number, ${\cal{R}}_0$, in the late-type stars using the mean field theory. It is widely believed now that the total X-ray luminosity, $L_X$, is a measure of stellar coronal activity. According to the observations, $L_X/L_* \propto {\cal{R}}_0^{-q}$, where $L_*$ is the bolometric luminosity, and $q$ is a constant. For ${\cal{R}}_0\gg1$, $2\leq q\leq3$, and for ${\cal{R}}_0\ll1$, $q=0$. Assuming that the X-ray flux depends on the stellar dynamo activity, they obtain an expression for the coronal luminosity in both ${\cal{R}}_0\gg 1$ and ${\cal{R}}_0\ll 1$ regimes. The calculated ratio is shown to have the same ${\cal{R}}_0$-dependence as the observations indicate. The authors construct an expression for $L_X/L_*$ as follows. Associate $L_X$ with stellar magnetic activity.  This is a well known theoretical model in which buoyancy causes the magnetic energy generated by a dynamo beneath the stellar surface to rise and dissipates into particles in the corona that radiate X-rays. Moreover, assume that the saturation strength of the large scale poloidal magnetic field is given by
\begin{equation}\label{Blackman-Thomas-1}
\frac{B_P^2}{8\pi}=\frac{l_{ed}}{L_1}f_h\rho v^2,
\end{equation}
where $l_{ed}$ and $v$ are, respectively, the eddy length scale and turbulent convective velocity, $\rho$ density, and 
$L_1=\xi h_p$ with $h_p$ being pressure scale height and $\xi \sim$ few. $f_h$ is the fractional magnetic helicity when the initial driver is kinetic helicity. 
Suppose that the toroidal field is linearly amplified by the shear during the buoyant loss time $\tau_b=L_1/u_b$ where $u_b$ is a typical buoyancy speed for the escaping structures. If $B\simeq B_{\phi}>B_P$, we get
\begin{equation}\label{Blackman-Thomas-2}
\frac{B^2}{8\pi}\sim \frac{B_{\phi}^2}{8\pi}\sim \frac{B_P^2}{8\pi}\left( \frac{\Omega}{\tau_b s} \right)^2,
\end{equation}
where $s$ is a constant that accounts for differential rotation. A strong surface
rotation indicates strong differential rotation and if a convective eddy is shredded by shear on a time scale $\tau_s<\tau_c$, where $\tau_c$ the convective turnover time, then the shorter shear time scale $\tau_s$ would become the relevant eddy correlation time such that $\tau_{ed}=\tau_s$. Now we assume that $\tau_s=s\tau_r$, with $\tau_r=2\pi/\Omega$, and also  
\begin{equation}\label{Blackman-Thomas-3}
u_b\simeq \frac{B_{\phi}^2 }{12\pi \rho v}.
\end{equation}
This expression has already obtained by several other authors (see \cite{BlackmanandThomas2015} and references therein). This leads to
\begin{equation}\label{Blackman-Thomas-4}
\tau_b\sim \frac{12\pi L_1 \rho v}{B_{\phi}^2}.
\end{equation}
Using this in eq.(\ref{Blackman-Thomas-2}) then gives
\begin{equation}\label{Blackman-Thomas-5}
\frac{B^2}{8\pi}\sim \frac{B_{\phi}^2}{8\pi}\sim \frac{ (12)^{2/3} }{8\pi^{1/3} } \left(\frac{\rho B_P\Omega v L_1}{s}\right)^{2/3},
\end{equation}
where $B_P$ is given by eq.(\ref{Blackman-Thomas-1}).

Estimate $L_X$ as
\begin{equation}\label{Blackman-Thomas-6}
L_X\sim L_{magnetic}\simeq \frac{B^2 u_b}{8\pi } \Theta r_c^2,
\end{equation}
where $r_c$ is the radius of the base of the convective zone, and $\Theta$ is the solid angle through which the field rises. Also,
\begin{equation}\label{Blackman-Thomas-7}
L_* \simeq 4\pi r_c^2 \rho v^3.
\end{equation}
Therefore, we find
\begin{equation}\label{Blackman-Thomas-8}
\frac{L_X}{L_*} \propto \left( \frac{L_1}{r_c}\right)^{2/3} \left(  \frac{s^{1/3} }{1+2\pi s {\cal{R}}_0 } \right)^2 \Theta,
\end{equation}

Assume
\begin{equation} \label{Blackman-Thomas-9}
\Theta =\Theta_0 \left(\frac{L_X/L_*}{6.6 \times 10^{-7} }\right)^{\lambda},
\end{equation}
with $\lambda=0$ and $\lambda=1/3$. For ${\cal{R}}_0\gg1$ regime, eq.(\ref{Blackman-Thomas-8}) gives ${L_X}/{L_*} \propto {\cal{R}}_0^{-2}$ with $\lambda=0$ while with $\lambda=1/3$, we get  ${L_X}/{L_*} \propto {\cal{R}}_0^{-3}$. Larger $\lambda$ makes $q>3$ whereas $2\leq q \leq 3$ is obtained from observations. Finally, note that ${L_X}/{L_*}$ is independent of ${\cal{R}}_0$ for ${\cal{R}}_0\ll1$.

However, one can question the validity of eq.(\ref{Blackman-Thomas-1}). There is mounting evidence that the ejection of magnetic helicity from the dynamo systems is critical for the dynamo efficiency. However, the assumption behind eq.(\ref{Blackman-Thomas-1}) is that there is no magnetic helicity flux within the system. Also, in eq.(\ref{Blackman-Thomas-6}), $B$ is assumed to be an average quantity distributed over the whole stellar surface with $\Theta$ representing a fraction while the real picture may not be that simple. The associated magnetic field is mostly considered on the stellar spots neglecting the role of $\Theta$. The functional form of $\Theta$ given by eq.(\ref{Blackman-Thomas-9}) is also questionable and requires more detailed considerations.

\section{Discussion}

Helicity is a scalar quantity which can be defined for any divergence free vector field. For example, because of the Gauss condition $\nabla.{\bf B}=0$, there is a vector potential $\bf A$ such that $\bf B=\nabla\times A$ and consequently the magnetic helicity is defined as the volume integral $\int_V {\bf A.B}d^3x$. This defines magnetic helicity density $J_M^0=\bf A.B$ which satisfies a continuity-like equation in ideal MHD; $\partial_t J_M^0+\nabla.{\bf J}_M=0$ where ${\bf J}_M=\phi{\bf B}+{\bf E\times A}$ is the helicity flux. In a relativistic setting, we can define magnetic helicity four-vector as $J_M^\mu=(J_M^0, {\bf J}_M)$ to write this conservation law in the compact form $\partial_\mu{J}_M^\mu=0$. This of course resembles the relativistic version of the continuity equation $\partial_\mu j^\mu=0$ for $j^\mu=(\rho, \bf j)$.

Electromagnetic helicities and their corresponding conservation laws arise very naturally from the Maxwell's equations. Magnetic helicity, however, plays a more crucial role than electric helicity partly because of the fact that in plasmas, e.g., in astrophysical fluids, magnetic field is stronger and more important than the electric field which vanishes because of the quasi-neutrality condition in neutral fluids. Unlike electric fields, therefore, large scale magnetic fields are observed in almost all cosmological objects such as stars, accretions disks and galaxies. The generation and evolution of these fields are studied in a vast literature which covers e.g., magnetic dynamo theories and magnetic reconnection models. Magnetic helicity is the only classical example of a Cherns-Simons symmetry; it is strictly conserved in ideal MHD and is conserved better than energy in resistive media, and its conservation law strongly constraints any theory related to magnetic field evolution. The most important example of such theories is perhaps magnetic dynamo theory. The conservation of magnetic helicity, surprisingly, is not considered in building many such models despite the fact that the role of helicity conservation was pointed out long time ago by Gruzinov and Diamond \cite{Gruzinovetal1994}, Vishniac and Cho \cite{VishniacandCho2001} and others. Any plausible dynamo theory must respect the magnetic helicity conservation to be self-consistent.

\appendix

\section{\textbf{Kinematic Dynamo Theory}}\label{s4.1}

In this appendix, we briefly revisit few important classic results in kinematic dynamo theory. In a dynamo, it is assumed that a weak seed field is amplified to energies comparable to the plasma turbulent kinetic energy. It is required also that the mechanism is such that maintains the dynamo action against dissipation. The kinematic theory is an approximation of dynamo problem in which the evolution of magnetic field is assumed to be governed by a velocity field. The basic assumption is that this velocity field is regarded as given. In the kinematic regime, it is assumed that strength of the generated magnetic field remains negligible, thus its back reaction on the flow can be neglected. However, the velocity field is usually, but not always, assumed to follow the Navier-Stokes equations. For non-turbulent kinematic dynamos, the velocity field is a smooth vector field which is often taken stationary. Turbulent kinematic dynamos can generate small scale or large scale magnetic fields, depending on which they are called small or large scale dynamos. Small scale dynamos generate magnetic fields whose gradient scale is of order the largest eddies in the turbulence whereas large scale dynamos generate fields with gradient scales comparable to the size of the system.

\iffalse

The onset of dynamo action is characterized by the magnetic Reynolds number, which is defined as $R_m=u_{rms}/\eta k_f$. Here, $u_{rms}$ is the root-mean-square velocity in the dynamo-active domain, $\eta$ is the magnetic diffusivity, and $k_f$ is the wavenumber corresponding to the energy-carrying scale of the flow.

\textbf{Small-scale dynamos} Small-scale dynamos can be studied analytically by solving evolution equations either for the correlation function $\langle BiBj\rangle$ or for the energy spectrum $E(k)$; see, e.g., the review by Brandenburg \& Subramanian (2005) with references to the original literature. In either approach the assumption of isotropy is usually invoked.

The critical magnetic Reynolds number for the onset of dynamo action is around $R_m=35$. The magnetic field has an approximate $k^{3/2}$ energy spectrum and is peaked at the resistive scale $\sim \eta/u_{rms}$. The growth rate scales with $R_m^{1/2}$.

At low magnetic Prandtl numbers $P_m=\nu/\eta$, the dynamo becomes harder to excite. This result does not, however, apply to dynamos with a mean flow (for example the Taylor-Green flow has a finite time average), or to flows with finite net helicity or anisotropy.

\textbf{Large-scale dynamos} For large-scale dynamos the magnetic energy grows at scales large compared with the scale of the turbulence. This requires that there is scale separation, i.e. that the domain size is large compared with the size of the turbulent eddies.
\fi

The evolution equation for a large scale field generated by a kinematic dynamo can be obtained by averaging the Maxwell's equations;

\begin{equation} \label{KT1}
\nabla\times \textbf{E}=- \frac{\partial \textbf{B}}{\partial t}, \; \nabla\times \textbf{B}=\frac{1}{c^2} \frac{\partial \textbf{E}}{\partial t}+\mu \textbf{J},
\end{equation}
\begin{equation} \label{KT2}
\nabla\textbf{.B}= 0, \; \nabla\textbf{.E}= \frac{\rho}{\epsilon}, 
\end{equation}
and also the Ohm's law;
\begin{equation}\label{KT3}
\textbf{J}=\sigma ( \textbf{E} +\textbf{V} \times \textbf{B}),
\end{equation}
This assumption that the velocity field of the fluid is always small allows us to neglect the term $({1}/{c^2}) {\partial \textbf{E}}/{\partial t}$. If we show the typical length and time scales of the system by $L$ and $T$, respectively, then

$$\frac{E}{L} \simeq \frac{B}{T}, \; \;  \frac{1}{c^2}|\frac{\partial \textbf{E}}{\partial t}| \simeq \frac{|\textbf{B}| L}{c^2T^2},\;\; |\nabla\times \textbf{B}| \simeq \frac{|\textbf{B}|}{L}.$$

Therefore, in systems where $L/T \ll c$ we have $({1}/{c^2}) {\partial \textbf{E}}/{\partial t} \ll |\nabla\times \textbf{B}|$. In astrophysical applications which we are interested in here, the length scale $L$ is the typical size of stars, planets etc, thus this assumption will obviously be held. We also need the induction equation given 

$$\frac{\partial \textbf{B}}{\partial t}=\eta {{\nabla}}^2 \textbf{B}+ \nabla\times ( \textbf{v} \times \textbf{B}).$$
This equation is the most important equation in dynamo theories. The kinematic dynamo theories consider the velocity field as a given function of position and time, $(\textbf{x},t)$. In a simple kinematic dynamo problem, we can suppose that velocity field is independent of time, and the problem is linear in terms of magnetic field. Thus, the problem can be solved looking for solutions of the form
\begin{equation}\label{KT4}
\textbf{B}=\textbf{B}_0(\textbf{x}) e^{at},\; \textbf{B}_0|_{\textbf{x} \rightarrow \infty}=0.
\end{equation}
The constant $a$ is, in general, a complex number, $a=\sigma +i \omega$. There can be an infinite number of eigenmodes, $\textbf{B}_0$, with the complex eigenvalue $a$. Here, $\sigma$ is the growth rate and $\omega$ is the frequency. In most cases, $\sigma$ is a negative number, so the diffusion term is dominant. This means that the solution is very oscillatory. If there are some modes with positive $\sigma$, there will be a dynamo. Growing modes, physically, mean that for an initial seed magnetic field, the system will amplify the magnetic field in a self-consistent way. However, for positive values of $\sigma$, the modes will not be growing forever. The reason is that we have to take into account the back-reaction of the generated magnetic field on the flow through the Lorentz force. This leads to a non-linear problem which is not considered in the framework of kinematic theories. For $\omega=0$, we have a steady growth which is responsible for steady dynamos. For, $\omega \neq 0$, which is the most common case, the growth is oscillatory and we will call them growing dynamo waves. Examples of kinematic dynamos are the Ponomarenko dynamo, and Roberts dynamo.

We can always use the decomposition into poloidal and toroidal components which was introduced in chapter 1. We know that a non-axisymmetric flow can create a non-axisymmetric magnetic field, while a non-axisymmetric magnetic field can also be generated by an axisymmetric flow. For axisymmetic magnetic field and velocity field, we may write a simpler decomposition for the magnetic field, 
\begin{equation}\label{KT5}
\begin{cases}
\textbf{B}=B_{\phi} \hat e_{\phi} +
\textbf{B}_P=B_{\phi}\; \hat e_{\phi}+\nabla\times (A \hat e_{\phi}), \\ 
    \textbf{v}=r \sin\theta\; \Omega \;\hat e_{\phi}+\textbf{v}_P=s\Omega \hat e_{\phi}+\nabla\times \frac{\psi}{s} \hat e_{\phi},
\end{cases}
\end{equation}
where $s=r \sin \theta$. Note that $r$ is the radius in the spherical geometry, however in the literature, it is also common to take $s=r$ which means $r$ is the radius in the cylindrical coordinate system. Now, the induction equation is written as two equations,
\begin{equation}\label{KT6}
\frac{\partial A}{\partial t} +\frac{1}{s} (\textbf{v}_P {.\nabla})(sA)=\eta ( {\nabla}^2-\frac{1}{s^2})A,
\end{equation}
\begin{equation}\label{KT7}
\frac{\partial B_{\phi}}{\partial t} +s (\textbf{v}_P {.\nabla})(\frac{B_{\phi}}{s})=\eta ( {\nabla}^2-\frac{1}{s^2})B_{\phi}+s\textbf{B}_P {.\nabla}\Omega.
\end{equation}
In both equations above, we have an advection term in the form $(\textbf{v}_P {.\nabla})$. This pushes the magnetic field around. On the other hand, the terms $({\nabla^2}-\frac{1}{s^2})$ are diffusion terms which are not capable of generating any magnetic field. The poloidal component has no source terms, and therefore, it will decay. If we want to keep it from decaying, we will need non-axisymmetric terms to sustain it. Also, note that the last term in the RHS of eq.(\ref{KT7}) will generate an azimuthal magnetic field as a result of stretching by rotation $\Omega$.

\textbf{\emph{Theorem 1.}}
 It is impossible to generate a two-dimensional dynamo (axisymmetric magnetic field).

This antidynamo theorem means that in Cartesian coordinates $(x,y,z)$, no dynamo action can maintain a magnetic field that is independent of $z$ and vanishes at infinity. For a two-dimensional field, the flow has to be two-dimensional. In Cartesian coordinates, the poloidal-toroidal decomposition is written as

\begin{equation}
{\bf B}=\nabla\times g \hat z)+\nabla\times(\nabla\times h\hat z)+b_x(z,t) \hat x+b_y \hat y.
\end{equation}

 If we use equations analogous to (\ref{KT6}), and (\ref{KT7}) by writing
\begin{equation}\label{KT8}
\begin{cases}
\textbf{B}=B_z \hat k +\textbf{B}_H=B_z \hat k + {\nabla} \times A \hat k, \\
\textbf{v}=v_z \hat k+\textbf{v}_H=v_z \hat k+\nabla\times \psi \hat k,
\end{cases}
\end{equation}
then, we find
\begin{equation}\label{KT9}
\frac{\partial A}{\partial t}+ (\textbf{v}_H {.\nabla})A=\eta {\nabla}^2A,
\end{equation}
\begin{equation}\label{KT10}
\frac{\partial B_z}{\partial t}+ (\textbf{v}_H {.\nabla})B_z=\eta {\nabla}^2B_z+\textbf{B}_H {.\nabla} v_z.
\end{equation}
Now, we multiply (\ref{KT9}) by $A$ and integrate it over the total volume of the system. This gives us
\begin{equation}\label{KT11}
\frac{\partial}{\partial t} \int_V \frac{1}{2} A^2 dV+ \int_V \frac{1}{2} {\nabla.} ({\textbf{v}}_H A^2) dV=-\eta \int_V ({\nabla}A)^2 dV
\end{equation}
We note that, in the above expression, the divergence term can be converted into a surface integral which is zero by assumption. Therefore, the integral of $A^2$ is equal to a definite negative integral which means it will decay. Given an enough time $\textbf{A}$ will decay to zero, and $\textbf{B}_H$ will vanish. As a result, there will be no source term in eq.(\ref{KT10}) which, applying the same line of reasoning, leads to $B=0$: there is no non-trivial two-dimensional magnetic field generated by a steady dynamo.

\textbf{\emph{Theorem 2.}} A planar velocity field, $(v_x,v_y,0)$ can not maintain a dynamo.

The $z$ component of the induction equation is

$$\frac{\partial B_z}{\partial t} +\textbf{v.}{\nabla}B_z=\eta {\nabla}^2B_z,$$

which can be multiplied by $B_z$ and integrated. The advection term leads to a surface integral which vanishes meaning that $B_z$ decays. We have $\nabla\textbf{.B}=0$, and if $B_z=0$, then $B_x=\partial_yA$, and $B_y=-\partial_x A$ for some $A$. Hence,

$$\partial_t (\partial^2_x+\partial^2_y)A+(\partial^2_x+\partial^2_y)(\textbf{v.}\nabla A)=\eta (\partial^2_x+\partial^2_y) {\nabla}^2A.$$

The Fourier transform of this expression gives the term $(k^2_x+k^2_y)$ which can be cancelled out. Thus, the above equation would be the same as eq.(\ref{KT9}). Then the latter expression, multiplied by $A$, shows that $A$ decays to zero.

\textbf{\emph{Theorem 3. (Cowling's theorem)}} It is not possible for an axisymmetric magnetic field which vanishes at infinity to be maintained by a dynamo action.

In fact, this theorem is the same as theorem 1 but in the polar coordinates. One can multiply eq.(\ref{KT6}) by the term $s^2A$, and integrate to find

$$
 \frac{\partial}{\partial t} \int_V \frac{1}{2} s^2 A^2 dV=-\eta \int_V |{\nabla}(sA)|^2 dV.$$

   This shows $A$ decays, and so $\textbf{B}_P=0$. In a similar way, we can multiply eq.(\ref{KT7}) by $B_{\phi}/s^2$ and integrate to obtain
$$  \frac{\partial}{\partial t} \int_V \frac{1}{2} s^{-2} B_{\phi}^2 dV=-\eta \int_V |{\nabla}(B_{\phi}/s)|^2 dV.$$
  This shows $B_{\phi}$ decays provided that it is not proportional to $s$. 

\textbf{\emph{Theorem 4.}} A purely toroidal flow, (i.e. $\textbf{v}=\nabla\times T_r$ with $v_r=0$) can not maintain a dynamo action.

This theorem is the same as theorem 2 in the polar coordinates. In this coordinate system, we can multiply 
$$\frac{\partial}{\partial t} (\textbf{r}\textbf{.B})+\textbf{v.}{\nabla}(\textbf{r.}\textbf{B})=\eta {\nabla}^2(\textbf{r.}\textbf{B})$$
by $(\textbf{r}\textbf{.B})$ and integrate which will prove the theorem.

\bibliographystyle{apsrev4-2}
\bibliography{IntroHelicity}

\end{document}